\documentclass[prd,preprintnumbers,aps,twocolumn,nofootinbib]{revtex4-2}
\usepackage{amsmath} 
\usepackage{amssymb}
\usepackage{graphicx,epsfig}
\usepackage[mathscr]{eucal}
\usepackage{color}
\usepackage{longtable}
\usepackage{natbib}
\usepackage[english]{babel}
\usepackage{tabularx}
\usepackage{setspace}
\usepackage{xspace}
\usepackage{multirow}
\usepackage{braket}
\renewcommand{\thefigure}{\arabic{figure}}

\newcommand\as{\alpha_{\mathrm{S}}}

\def\to{\rightarrow} 
\def\nn{\nonumber}

\def\ktness{k_T^{\rm ness}}

\usepackage{scalefnt,pstricks}

\def\rcut{r_{\rm cut}}

\newcommand\MCFM{{\sc MCFM}\xspace}
\newcommand\POWHEG{{\sc POWHEG}\xspace}

\newcommand\OpenLoops{{\sc OpenLoops}\xspace}

\usepackage{array}
\newcolumntype{L}[1]{>{\raggedright\let\newline\\\arraybackslash\hspace{0pt}}m{#1}}
\newcolumntype{C}[1]{>{\centering\let\newline\\\arraybackslash\hspace{0pt}}m{#1}}
\newcolumntype{R}[1]{>{\raggedleft\let\newline\\\arraybackslash\hspace{0pt}}m{#1}}

\usepackage[colorlinks=true,allcolors={blue!70!black}]{hyperref}

\begin{document} 
\preprint{ZU-TH 04/22}

\title{Effective transverse momentum in multiple jet production at hadron colliders}

\author{Luca Buonocore, Massimiliano Grazzini, J\"urg Haag, Luca Rottoli and Chiara Savoini}

\affiliation{Physik Institut, Universit\"at Z\"urich, CH-8057 Z\"urich, Switzerland}

\begin{abstract}
We consider the class of inclusive hadron collider processes in which several energetic jets are produced,   
possibly accompanied by colourless particles (such as Higgs boson(s), vector boson(s) with their leptonic decays, and so forth).
  We propose a new variable that smoothly captures the 
  $N+1$ to $N$-jet transition.
  This variable, that we dub $\ktness$, represents an effective transverse momentum controlling the singularities of the $N+1$-jet cross section when the additional jet is unresolved.
  The $\ktness$ variable offers novel opportunities to perform higher-order calculations in Quantum Chromodynamics (QCD) by using non-local subtraction schemes.
  We study the singular behavior of the $N+1$-jet cross section as $\ktness\to 0$ and, as a phenomenological application,  we use the ensuing results to evaluate next-to-leading order corrections to $H$+jet and $Z$+2 jet production at the LHC. 
  We show that $\ktness$ performs extremely well as a resolution variable and appears to be very stable with respect to hadronization and multiple-parton interactions.  

\end{abstract}

\maketitle

\section{Introduction}

Most of the new-physics searches and precision studies of the Standard Model at the Large Hadron Collider (LHC) are carried out by identifying events with a definite number of energetic leptons, photons and jets.
Jets are collimated bunches of hadrons that represent the fingerprints of the high-energy partons (quarks and gluons) produced in the hard-scattering interaction.
A precise description of  jet processes requires observables capable of efficiently capturing
the dynamics of the energy flow in hadronic final states.
A classic example is provided by dimensionless {\it shape variables} in $e^+ e^-$ collisions~\cite{Dasgupta:2003iq} and their generalisation to proton-proton collisions~\cite{Banfi:2004nk,Banfi:2010xy}.
These variables are sensitive to different aspects of the theoretical description of the underlying hard-scattering process and are designed to measure the deviation from the leading order (LO) energy flow, which characterises the bulk of the events.

For processes with $N$ jets at the Born level, observables describing the $N+1\to N$ jet transition are particularly useful to veto additional jets, for instance to discriminate signal over backgrounds.
When no jets are produced in the final state and only a colourless system is tagged, a prominent example of a dimensionful variable which inclusively describes the initial-state radiation is given by the transverse momentum of the colourless system ($q_T$).
Another commonly used variable is ($N$-)jettiness \cite{Stewart:2010tn}, $\tau_N$, which is defined on events containing at least $N$ hard jets.
Requiring $\tau_N\ll 1$ constrains the radiation outside the signal (and beam) jets,
effectively providing an inclusive way to veto additional jets.
These resolution variables have been used to formulate non-local subtraction methods for QCD calculations at next-to-next-to-leading order (NNLO) and beyond \cite{Catani:2007vq,Gaunt:2015pea}.  Both $q_T$ and $\tau_N$ are also used as resolution variables in the matching of NNLO computations to parton shower simulations~\cite{Alioli:2013hqa,Hoche:2014dla,Monni:2019whf,Mazzitelli:2020jio,Alioli:2021qbf}.

In the context of non-local subtraction schemes, the efficiency of the calculation is subject to the size of the (missing) power-corrections, which in general depends on the choice of the resolution variable.
For multijet processes, jettiness is the only viable variable proposed to date, and it has been successfully used to compute NNLO corrections to several colour-singlet processes~\cite{Gaunt:2015pea,Boughezal:2016wmq,Campbell:2016yrh,Heinrich:2017bvg,Campbell:2017aul} and to the production of a vector or Higgs boson in association with a jet~\cite{Boughezal:2015aha,Boughezal:2015ded,Boughezal:2016dtm}. Nonetheless, it is well known that $\tau_N$ is affected by large power corrections. Indeed, even in the case of the production of a colourless final state,
the power suppressed contributions are linear and logarithmically enhanced already at next-to-leading order (NLO), see e.g.~Refs.~\cite{Moult:2016fqy,Boughezal:2016zws,Boughezal:2018mvf,Ebert:2018lzn,Campbell:2019gmd}.
On the other hand, non-local subtraction methods based on $q_T$ are subject to milder power corrections, which can even be quadratic~\cite{Grazzini:2016ctr,Ebert:2018gsn,Cieri:2019tfv} in the absence of cuts \cite{Grazzini:2017mhc,Buonocore:2019puv,Ebert:2019zkb,Catani:2019iny}.

An important probe of QCD dynamics in the infrared region is obtained when fixed-order perturbative predictions are supplemented with the all-order resummation of soft and collinear emissions and eventually compared to experimental data.
This comparison largely depends on low-energy physics phenomena, such as hadronization or multiple-parton interactions (MPI), which must be properly included to realistically simulate collider events.
In particular, the constraining power of the observable is substantially diluted when these effects, which go beyond a purely perturbative description, become dominant.
While hadronization corrections and MPI effects are mild in the case of $q_T$, they are
particularly significant in the case of observables like $N$-jettiness~\cite{Gaunt:2014ska}.

In this work we introduce a new global dimensionful observable, that we call $\ktness$, to describe the $N+1\to N$ jet transition. This variable, which takes its name
from the $k_T$ jet clustering algorithm \cite{Catani:1993hr,Ellis:1993tq}, represents an effective transverse momentum
describing the limit in which the additional jet is unresolved. When the unresolved radiation is close to the colliding beams, the variable coincides with the transverse momentum
of the final-state system. When the unresolved radiation is emitted close to one of the final-state jets, the variable describes the relative transverse momentum with respect to the
jet direction. As we will show, $\ktness$ offers a number of attractive features.
It is affected by relatively mild power corrections, thereby allowing us to efficiently compute NLO corrections to processes with a vector or Higgs boson plus one or more jets.
We also show that the new variable is very stable with respect to hadronization and MPI. Due to these appealing properties, $\ktness$ might prove useful as
a resolution variable in higher order computations and in their matching to parton shower simulations, as well as in studies of non-perturbative physics associated to the underlying event in proton-proton collisions.

The paper is organized as follows.
In Section~\ref{sec:def} we define $\ktness$ and we discuss the formulation of a subtraction scheme based on this observable.
We present numerical results for Higgs$+$jet production and for $Z+2$ jet production at NLO in Section~\ref{sec:results}.
We also study the impact of MPI effects and hadronization on the $\ktness$ observable in $Z+1$ jet production. 
Finally, we summarize our findings and discuss future prospects in Section~\ref{sec:conclusion}.
Further details on the perturbative ingredients needed at NLO are given in Appendix~\ref{sec:appendix}.

\section{Definition and NLO implementation}\label{sec:def}

We consider the inclusive hard-scattering process
\begin{equation}
  \label{eq:process}
  h_1(P_1)+h_2(P_2)\! \to\! j(p_1)+j(p_2)+...+j(p_N)+F(p_F)+X
\end{equation}
where the collision of the two hadrons $h_1$ and $h_2$ with momenta $P_1$ and $P_2$ produces $N$ final-state hard jets with momenta $p_1$, $p_2$...$p_N$,
possibly accompanied by a generic colourless system $F$ with total momentum $p_F$.
The QCD radiative corrections to the process in Eq.~(\ref{eq:process}) receive contributions from final states including up to $N+k$ partons, where $k$ is the order of the computation.
The dimensionful quantity $(N$-)$k_T^{\rm ness}$ for an event featuring $N+m$ partons (with $1\leq m\leq k$) is defined
by using the {\it exclusive} $k_T$ clustering algorithm \cite{Catani:1993hr,Ellis:1993tq}. We first introduce the distances
  \begin{equation}
    d_{ij}={\rm min}(p_{Ti},p_{Tj})\Delta R_{ij}/D, \qquad d_{iB}=p_{Ti}
    \label{eq:distance}
    \end{equation}
  where $D$ is a parameter of order unity, $i,j=1,2...,N+m$ and $\Delta R^2_{ij}=(y_i-y_j)^2+(\phi_i-\phi_j)^2$ is the standard separation in rapidity ($y$) and azimuth ($\phi$) between the (pseudo)particles $i$ and $j$. The quantity $d_{iB}$ is the {\it particle-beam} distance, given by the transverse momentum $p_{Ti}$ \footnote{To be precise, Refs.~\cite{Catani:1993hr,Ellis:1993tq} use pseudorapidities in the definition of $d_{ij}$. Here we use rapidities as in the implementation of the $k_T$ algorithm in the \textsc{Fastjet} code~\cite{Cacciari:2011ma}.}. 
  The $k_T^{\rm ness}$ variable is defined via a recursive procedure through which close-by particles are combined with each other or with the beam until $N$+1 jets remain.
The procedure goes as follows:
\begin{enumerate}
\item Compute the minimum of the $d_{ij}$ and the $d_{iB}$.
  If there are at least $N$+2 final-state pseudoparticles perform step 2. If there are only $N$+1 pseudoparticles perform step 3.
\item If the minimum is one of the $d_{iB}$, then recombine $i$ with the beam and remove it from the list of pseudoparticles. 
  The recombination is done starting from a recoil momentum initialised to $p_{\rm rec}=0$ at the beginning of the procedure and collecting the recoiled momenta with $p_{\rm rec} \rightarrow p_{\rm rec}+p_i$. If the minimum is a $d_{ij}$ then replace the pseudoparticles $i$ and $j$ with a new pseudoparticle with momentum $p_i+p_j$. Go back to step 1 with a configuration which has one pseudoparticle less.  
\item
When $N+1$ pseudoparticles are left, if the minimum is one of the $d_{iB}$ add the recoil to $p_i$ through $p_i \rightarrow p_{\rm rec}+p_i$ and set $k_T^{\rm ness}=p_{T,i}$. If instead the minimum is a $d_{ij}$ then set $k_T^{\rm ness}={\rm min}(d_{ij})$.
\end{enumerate}
\vspace*{.1cm}
To the best of our knowledge, $\ktness$ has not been considered before in the literature.
The variable depends on the parameter $D$ entering the distance in Eq.~(\ref{eq:distance}).
We note that other prescriptions to treat the recoil (for instance, by neglecting it in step 3) are in principle possible,
and the differences start to appear from NNLO (i.e. $k=2$).

We have computed the singular behavior of the cross section for the production of a colourless system accompanied by an arbitrary number of jets at NLO as \mbox{$\ktness\to 0$}.
The computation starts by organizing the terms relevant in each singular region and removing the double counting, similarly to what is done in Refs.~\cite{Catani:2014qha,Buonocore:2021akg}.
The terms containing initial-state collinear singularities produce the so called {\it beam} functions,
while those containing final-state collinear singularities give rise to the {\it jet} functions.
The remaining contributions, describing soft radiation at large angles, produce the so called {\it soft} function.
The soft and collinear singularities, regulated by working in $d=4-2\epsilon$ dimensions, cancel out with those of the virtual contribution, to obtain a finite cross section.
The results of our computation are used to construct a subtraction formula for the partonic cross section $d{\hat \sigma}_{\rm NLO}^{\rm F+N\,jets+X}$ as follows
\begin{align}
    \label{eq:main}
  d{\hat \sigma}_{\rm NLO}^{\rm F+N\,jets+X}&={\cal H}^{\rm F+N\,jets}_{\rm NLO}\otimes d{\hat \sigma}_{\rm LO}^{\rm F+N\,jets}+\nn\\
  &+\left[d{\hat \sigma}_{\rm LO}^{\rm F+(N+1)\,jets}-d{\hat \sigma}^{\rm CT,F+Njets}_{\rm NLO}\right]\, .
\end{align}
The real contribution $d\hat \sigma_{\rm LO}^{\rm F+(N+1)jets}$ is obtained by integrating the tree-level matrix elements with one additional parton and is divergent in the limit $\ktness\to 0$.
The {\it counterterm} $d\hat \sigma^{\rm CT,F+Njets}_{\rm NLO}$ is constructed by combining the singular contributions discussed above and matches the real contribution in the $\ktness\to 0$ limit. Its explicit expression in the partonic channel $ab$ reads
\begin{widetext}
\begin{align}
  d&{\hat \sigma}^{\rm CT,F+Njets}_{{\rm NLO}\,  ab}  =  
  \frac{\as}{\pi}   \frac{d \ktness}{\ktness}\!  \Bigg\{ \Bigg[ \ln \frac{Q^2}{(\ktness)^2} \sum_{\alpha} C_{\alpha}-\sum_\alpha \gamma_\alpha
 - \sum_i C_{i} \ln \left(D^2\right)
 - \sum_{\alpha\neq\beta} \langle\mathbf{T}_\alpha \cdot \mathbf{T}_{\beta}\rangle \ln \left(\frac{2p_\alpha\cdot p_\beta}{Q^2} \right)\Bigg]\times\nonumber\\
&\times  \delta_{ac}\delta_{bd}\delta(1-z_1)\delta(1-z_2)
+ 2\delta(1-z_2)\delta_{bd}P^{(1)}_{ca}(z_1)+2\delta(1-z_1)\delta_{ac}P^{(1)}_{db}(z_2) \Bigg\} \otimes d{\hat \sigma}^{\rm F+N\, jets}_{{\rm LO}\, cd},
\end{align}
\end{widetext}
where $\gamma_q=3C_F/2$, $\gamma_g=(11C_A-2n_F)/6$, $C_F=4/3$ and $C_A=3$ are the QCD colour factors with $n_F$ the number of active flavours and $D$ is the parameter entering the
definition of $\ktness$ (see Eq.~(\ref{eq:distance})).
The index $i$ labels the final-state partons with colour charges $\mathbf{T}_i$ ($\mathbf{T}^2_i=C_i$) and momenta $p_i$ ($\sum_i p_i=q$, $Q^2=q^2$), while $\alpha$ and $\beta$ label initial and final-state partons.
The symbol $\langle\mathbf{T}_\alpha \cdot \mathbf{T}_{\beta}\rangle=\langle{\cal M}_{cd\to {\rm F+N\,jets}}|\mathbf{T}_\alpha \cdot \mathbf{T}_{\beta}|{\cal M}_{cd\to {\rm F+N\,jets}}\rangle/|{\cal M}_{cd\to {\rm F+N\,jets}}|^2$ is the normalised colour-correlated tree level matrix element for the partonic process contributing to $d{\hat \sigma}^{\rm F+N\, jets}_{{\rm LO}\, cd}$, and a sum over all the possible final-state parton flavours is understood.
The functions $P^{(1)}_{ab}(z)$ are the LO Altarelli-Parisi kernels (in $\as/\pi$ normalisation)~\cite{Altarelli:1977zs,Dokshitzer:1977sg,Gribov:1972ri}, and
the symbol $\otimes$ denotes the convolutions with respect to the longitudinal-momentum fractions $z_1$ and $z_2$ of the colliding partons.
The square bracket in Eq.~(\ref{eq:main}) is evaluated by requiring $\ktness/M>\rcut$, where $M \sim Q$ is a hard scale which can be chosen
depending on the specific process under consideration.
The first term on the right-hand side of Eq.~(\ref{eq:main}) is obtained by convoluting
the LO cross section $d\hat \sigma_{\rm LO}^{\rm F+N\,jets}$
with the perturbative function ${\cal H}_{\rm NLO}^{\rm F+N\,jets}$. The latter
contains the virtual correction after subtraction of the infrared singularities, additional finite contributions of collinear origin (beam and jet functions)
and of soft origin (soft function). More details on the evaluation of ${\cal H}_{\rm NLO}^{\rm F+N\,jets}$ can be found in Appendix~\ref{sec:appendix}.
The physical cross section is formally obtained by taking the limit $\rcut\to 0$ in Eq.~(\ref{eq:main}).

We have implemented Eq.~(\ref{eq:main}) to evaluate $H+{\rm jet}$ (in the limit of an infinite top-quark mass) and $Z+2$~jet production at the LHC.
The real contribution is evaluated with \MCFM~\cite{Campbell:2019dru}, while the subtraction counterterm
$d{\hat \sigma}^{\rm CT,F+Njets}_{\rm NLO}$ and the ${\cal H}^{\rm F+N\,jets}_{\rm NLO}\otimes d{\hat \sigma}_{\rm LO}^{\rm F+N\,jets}$
contribution are computed with a dedicated implementation.
In particular, for $H+$ jet production the required tree-level and one-loop amplitudes are still obtained from \MCFM,
while for $Z+2$~jet all the (colour-correlated) amplitudes are evaluated with \OpenLoops \cite{Cascioli:2011va, Buccioni:2017yxi, Buccioni:2019sur}.

\section{Results}\label{sec:results}

\begin{figure}[t]
  \begin{center}
  \includegraphics[width=\columnwidth]{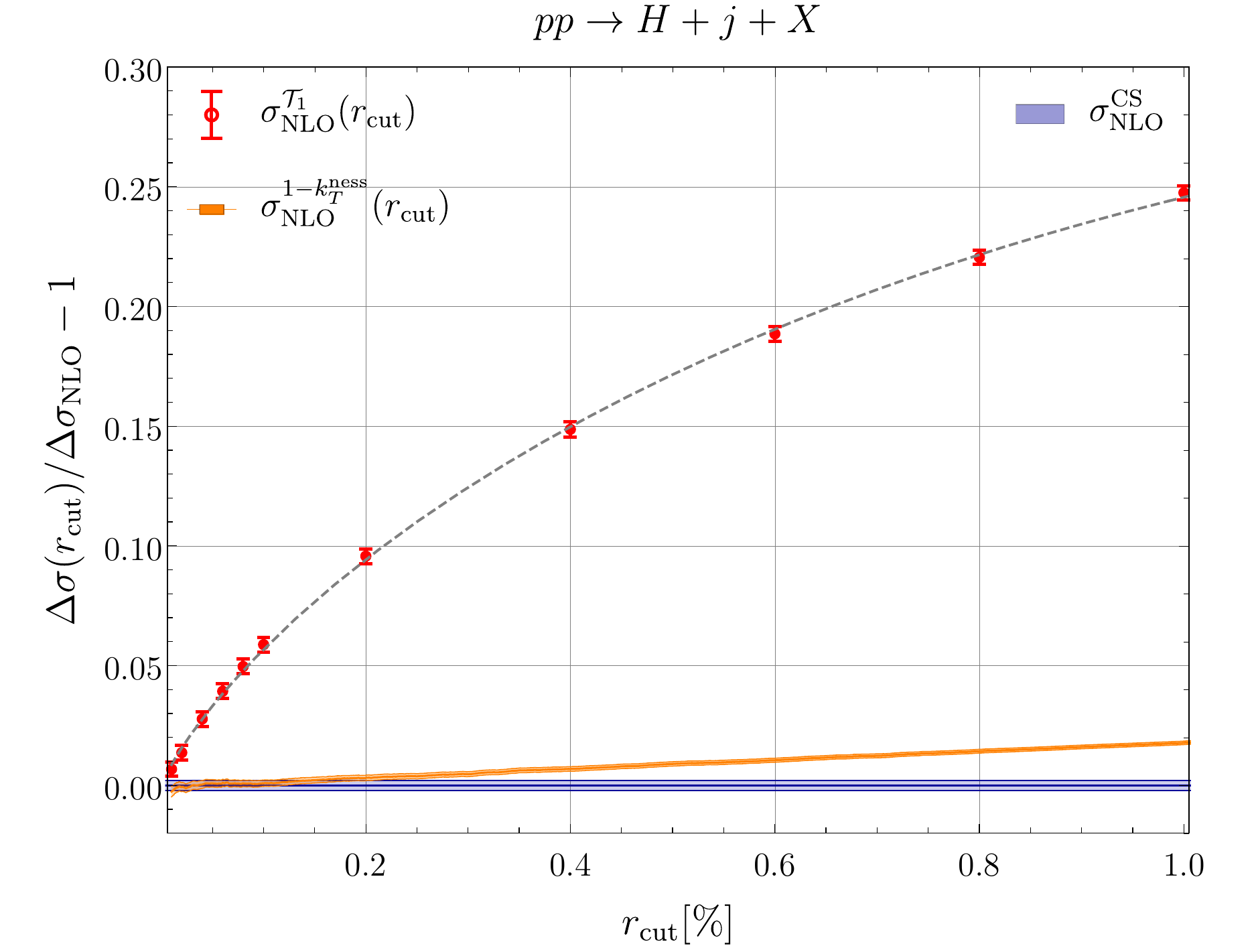}\\[2ex]
\end{center}
\vspace{-2ex}
\caption{\label{fig:Hjet}
The NLO correction $\Delta\sigma$ for the $H+{\rm jet}$ cross section computed with 1-jettiness (red points) and 1-$\ktness$ (orange). The $\rcut$ dependence is compared to the ($\rcut$ independent) result obtained with dipole subtraction using \MCFM (blue).}
\end{figure}

We consider proton-proton collisions at the LHC at a centre-of-mass energy of 13 TeV. 
We use the \texttt{NNPDF31\_nlo\_as\_0118} parton distribution functions~\cite{NNPDF:2017mvq} with $\as(m_Z) = 0.118$ through the \textsc{Lhapdf} interface~\cite{Buckley:2014ana}.
As for the electroweak couplings we use the $G_\mu$ scheme with $G_F=1.16639\times 10^{-5}$ GeV$^{-2}$, $m_W=80.385$ GeV, $m_Z=91.1876$ GeV, $\Gamma_Z=2.4952$ GeV.
We define jets via the anti-$k_T$ algorithm~\cite{Cacciari:2008gp} with $R=0.4$.

We start the presentation of our results with $H+{\rm jet}$ production.
We compute the corresponding cross section through Eq.~(\ref{eq:main}) by setting the parameter $D=1$ and
requiring the leading jet to have $p_T^{j} > 30$~GeV.
The factorization and renormalization scales $\mu_F$ and $\mu_R$ are set to the Higgs boson mass $m_H=125$~GeV.
In order to compare our results to those that can be obtained with jettiness subtraction,
we have implemented the corresponding calculation in a modified version of the \MCFM code~\cite{Campbell:2019dru}, which we have benchmarked against the numerical results of Ref.~\cite{Campbell:2019gmd}.
The $1$-jettiness variable is defined as
\begin{equation}
  \mathcal T_1 = \sum_i \min_l \left\{ \frac{2 q_l \cdot p_i}{Q_l} \right\},
  \label{eq:1jettiness}
\end{equation}
where $q_l$ ($l=1,2,3$) are the momenta of the initial-state partons and of the hardest jet present in the event and the sum over $i$ runs over the final-state parton momenta $p_i$.
Following Ref.~\cite{Campbell:2019gmd} we compute $\mathcal T_1$ in the hadronic centre-of-mass frame and we set the normalization factors $Q_l=2 E_l$.
To compare the results obtained with a $1$-jettiness cut to those obtained using $1$-$k_T^{\rm ness}$ we define the minimum $\rcut$ on the
dimensionless variable $r={\cal T}_1/\sqrt{m_H^2 + (p_T^j)^2}$ ($r=\ktness/\sqrt{m_H^2 + (p_T^j)^2}$).

\begin{figure}[t]
  \begin{center}
  \includegraphics[width=0.95\columnwidth]{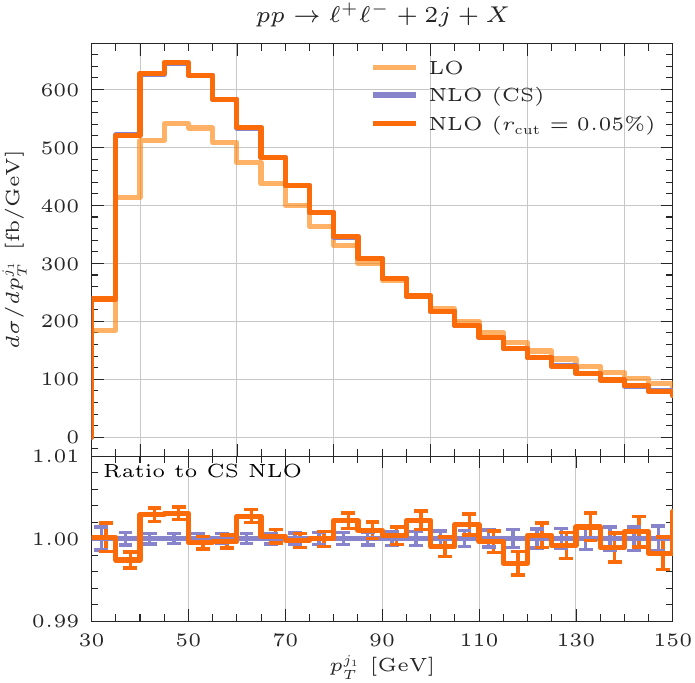}\\
  \hspace*{-0.25cm}
  \includegraphics[width=0.95\columnwidth]{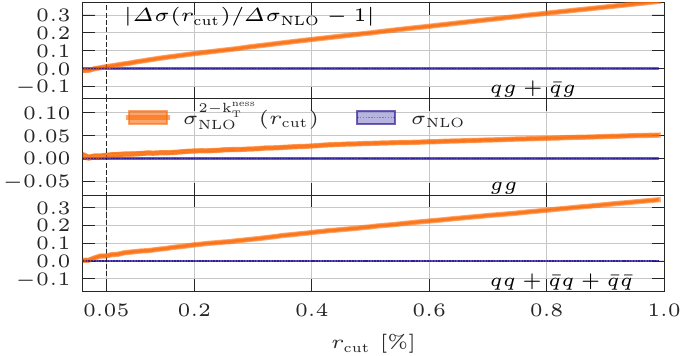}
\end{center}
\vspace{-2ex}
\caption{\label{fig:Z2jet}
  $Z+2$~jet production at NLO: $\ktness$-subtraction against CS. The $p_T$ distribution of the leading jet (upper and central panels) at LO (yellow) and NLO (orange: $\ktness$, blue: CS). NLO corrections $\Delta\sigma$ as a function of $\rcut$ in the three partonic channels (lower panels).}
\end{figure}

In Fig.~\ref{fig:Hjet} we study the behavior of the NLO correction $\Delta\sigma$ as a function of $\rcut$ for the jettiness and $\ktness$ calculations, normalised to the result obtained with Catani-Seymour (CS) dipole subtraction \cite{Catani:1996jh,Catani:1996vz}
by using \MCFM (which is independent of $\rcut$). Both jettiness and $\ktness$ results nicely converge to the expected result but the $\rcut$ dependence is very different for the two calculations.
The $\rcut$ dependence in the case of jettiness is rather strong: At $\rcut=1\%$ the difference with the exact result is about $25\%$ of the computed correction.
The observed $\rcut$ dependence is consistent with a logarithmically enhanced linear behavior.
We note that performing the computation in other frames may improve the convergence~\cite{Campbell:2019gmd}, but the functional behavior remains the same.  
By contrast, in the case of $\ktness$ the dependence is rather mild. At $\rcut=1\%$ the difference with the exact result is only about $3\%$ of the computed correction.
The $\rcut$ dependence is consistent with a purely linear behavior (i.e. without logarithmic enhancements).

We now move to consider $Z+2$~jet production. We compute the cross section to obtain a dilepton pair in the invariant-mass range 66 GeV $\leq m_{\ell\ell}\leq 116$ GeV together with (at least) 2 jets with
$p_T>30$ GeV and pseudorapidity $|\eta|<4.5$. The leptons have $p_T\geq 20$ GeV and pseudorapidity $|\eta_\ell|\leq 2.5$.
The minimum separation between the leptons is $\Delta R_{\ell\ell}>0.2$ while leptons and jets have $\Delta R_{\ell j}>0.5$.
The factorization and renormalization scales are set to the $Z$ boson mass $m_Z$.
Our calculation is carried out by using the transverse mass of the dilepton system as a hard scale $M$ to define $\rcut$ and
the parameter $D$ is set to $D=0.1$ in this case. 
We have checked that similar results can be obtained by choosing different values of $D$.
We compare our results with those obtained using the implementation of $Z+2$~jet production of~\cite{Kallweit:2015dum}, which is based on CS subtraction.
In Fig.~\ref{fig:Z2jet} (upper panel) we show the $p_T$ distribution of the hardest jet at LO and NLO, computed with $\ktness$ subtraction (using $\rcut=0.05 \%$) and with CS.
The central panel shows the relative difference between the two calculations. We observe an excellent agreement between the two results at the few permille level.
The three lower panels display the NLO correction $\Delta\sigma$ as a function of $\rcut$ in the \mbox{(anti-)quark}-gluon, gluon-gluon and \mbox{(anti-)quark}-\mbox{(anti-)quark} partonic channels compared to the corresponding result obtained with CS. The results nicely converge to the CS values in all the channels, and also in this case the $\rcut$ dependence is linear.

\begin{figure}[t]
  \begin{center}
   \begin{tabular}{cc} 
  \includegraphics[width=0.523\columnwidth]{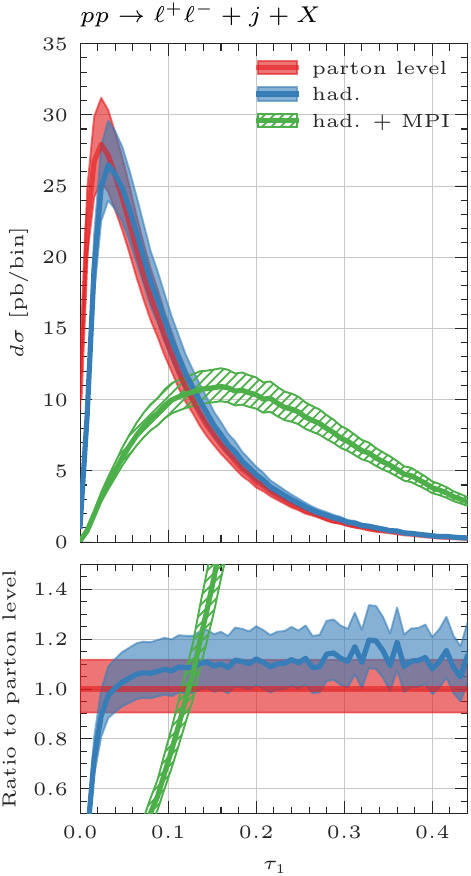}&
  \hspace*{-0.28cm}
  \raisebox{-0.008\height}{\includegraphics[width=0.465\columnwidth]{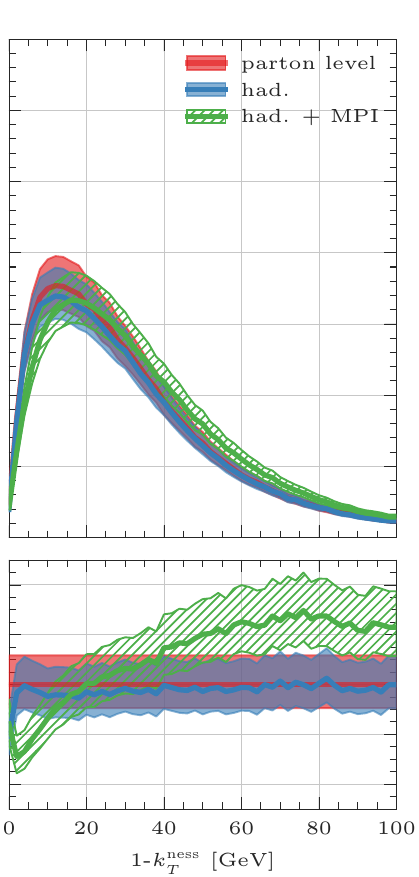}}\\[2ex]
  \end{tabular}
\end{center}
\vspace{-2ex}
\caption{\label{fig:mc}
$Z+{\rm jet}$ at LO + parton shower: $\tau_1$ (left panel) and 1-$\ktness$ (right panel) spectra at the parton level (red) and including hadronization (blue) or hadronization and MPI (green).}
\end{figure}

Finally, in view of potential applications of $\ktness$ as a probe of jet production in hadron collisions, 
we study the stability of our new variable under hadronization and MPI. We have generated a sample of LO events for $Z+{\rm jet}$ with the \POWHEG Monte Carlo event generator~\cite{Nason:2004rx,Frixione:2007vw,Alioli:2010qp} and showered them with \textsc{Pythia8}~\cite{Sjostrand:2014zea} using the A14 tune~\cite{TheATLAScollaboration:2014rfk}.
We use the same setup as for $H+{\rm jet}$, now setting $\mu_R=\mu_F=m_Z$ and adding an additional requirement on the leading jet rapidity $|y^{j_1}| < 2.5$. 
We define the (dimensionless) $1$-jettiness event shape $\tau_1$ as in Ref.~\cite{Stewart:2010tn}; the jet axis coincides with the direction of the leading jet reconstructed using the \textsc{Fastjet} code~\cite{Cacciari:2011ma} (which we also use in each step of the $k_T$-clustering algorithm used to compute $\ktness$). This simply corresponds to choosing $Q_l$ as the partonic centre-of-mass energy $Q$ in Eq.~(\ref{eq:1jettiness}) and to defining $\tau_1={\mathcal T}_1 /Q$.
Our results are shown in Fig.~\ref{fig:mc}. The left panel shows the $1$-jettiness distribution while the right panel depicts the $1$-$\ktness$ result. The result obtained at parton level (red) is compared with the result including hadronization corrections (blue) and further adding MPI (green).
The bands are obtained by varying $\mu_F$ and $\mu_R$ by a factor of 2 around their central value with the constraint $1/2<\mu_F/\mu_R<2$.
The $1$-jettiness distribution has a Sudakov peak at $\tau_1\sim 0.02$. The hadronization corrections are relatively large in the region of the peak, and remain of the order of $10\%$ as $\tau_1$ increases.
The inclusion of MPI drastically changes the shape of the distribution, the peak moving to $\tau_1\sim 0.15$.
The $1$-$\ktness$ distribution displays a Sudakov peak at $1$-$\ktness\sim 15$ GeV, similarly to what we would expect from a transverse-momentum distribution of a colourless system produced by gluon fusion.
The position of the peak and the shape of the distribution remain rather stable when hadronization is included, while the inclusion of MPI makes the distribution somewhat harder.
Comparing the left and right panels of Fig.~\ref{fig:mc} we clearly see that the $\ktness$ distribution is significantly more stable against the inclusion of hadronization and MPI, and it is therefore a good candidate for
QCD studies in multijet production at hadron colliders.
This could have maybe been expected, since variables based on transverse momenta are known \cite{Banfi:2010xy}
  to be mildly sensitive to hadronisation and underlying event.

\section{Summary and Outlook}\label{sec:conclusion}
In this work we have introduced the new variable $\ktness$ to describe multiple jet production in hadronic collisions.
The variable represents an effective transverse momentum controlling the singularities of the $N+1$-jet cross section when the additional jet is unresolved.
The $\ktness$ variable can be used to perform higher-order QCD calculations by using a non-local subtraction scheme, analogously to
what is done for $q_T$ and jettiness.
  We have computed the singular limit of the $N+1$-jet cross section as $\ktness\to 0$
  and we have used the results to evaluate NLO corrections to $H+{\rm jet}$ and $Z+2$ jet production at the LHC,
  finding complete agreement with results obtained with standard NLO tools.
  Compared to jettiness, power corrections are under much better control and are linear, without logarithmic enhancements.
  This scaling behaviour, which is due to the fact that $\ktness$ is an effective transverse momentum,
    is expected to be a general property holding for an arbitrary number of jets, since no additional perturbative ingredients appear beyond the two jet case~\footnote{We note that such linear scaling behaviour has a {\it dynamical} origin, and, therefore, cannot be captured through recoil effects as in Refs.~\cite{Buonocore:2021tke,Camarda:2021jsw}}.
  The extension of our calculations to NNLO will require a significant amount of conceptual and technical work, but
  our results suggest that, once the missing perturbative ingredients are available,
  an NNLO scheme based on $\ktness$ could be constructed and implemented in a relatively simple way, as done for $q_T$ in the case of colourless final states and heavy quarks \cite{Grazzini:2017mhc,Catani:2019hip,Catani:2020kkl}.
 Additionally, the perturbative coefficients entering the NLO and NNLO calculations are necessary ingredients to study the all-order structure of $\ktness$ at next-to-leading logarithmic  accuracy and beyond.
  We have also shown that $\ktness$ appears to be very stable with respect to hadronization and multiple-parton interactions, thereby offering new opportunities for
  QCD studies in multijet production at hadron colliders.
  Being an effective transverse momentum, we also expect that $\ktness$ could be used as a resolution variable when matching NNLO computations to $k_T$-ordered parton showers for processes with one or more jets at the Born level. We look forward to further studies in these directions.

\begin{acknowledgements}

\noindent This work is supported in part by the Swiss National Science Foundation~(SNF) under contracts 200020$\_$188464 and PZ00P2$\_$201878 and by the UZH Forschungskredit Grants K-72324-03 and FK-21-102. We are grateful to Stefano Catani, Stefan Kallweit, Pier Monni, Gavin Salam, Gregory Soyez, Marius Wiesemann for comments on the manuscript and several discussions.
	
\end{acknowledgements}

\newpage

\appendix
\section{Jet and soft functions}\label{sec:appendix}
\noindent In the following we discuss the evaluation of the perturbative coefficient ${\cal H}_{\rm NLO}^{\rm F+N\,jets}$ controlling the first term
in the subtraction formula Eq.~(3).
To set up our notation we consider the $N$-jet partonic process,
\begin{equation} 
  \label{eq:pprocess}
    c(p_c) + d(p_d) \rightarrow j(p_1) + j(p_2) + ... + j(p_{N}) + F(p_F)
\end{equation}
which contributes to Eq.~(1) at Born level.
The perturbative coefficient ${\cal H}_{cd;ab}^{\rm F+N\,jets}$ to be convoluted with the partonic cross section $d{\hat \sigma}_{{\rm LO}~cd}^{\rm F+N~jets}$ can be written as
\begin{equation}
    {\cal H}^{\rm F+N\,jets}_{cd;ab}= ({\bf H S})_{cd} C_{ca}C_{db}\prod_{i=1,...,N}\,J_i
\end{equation}
where $C_{ca}$ and $C_{db}$ are the customary collinear functions appearing in the $q_T$ subtraction formalism, $J_i$ are the jet functions
describing collinear radiation to each of the final-state parton,
and an appropriate sum over final-state parton flavors is understood. The explicit expressions of the jet functions read
\begin{align}
    J^f_i = 
    \begin{cases}
        1 + \frac{\as(\mu_R)}{\pi} \biggl\{ C_A\biggl[ \frac{131}{72} -\frac{\pi^2}{4} -\frac{11}{6}\ln(2D) \\ ~~~~~~~~~~~~~~~~~~-\ln(D)\ln\biggl(\frac{Q^2}{4E^2_i} \biggr) -\ln^2(D)\biggr)\biggr]\nonumber\\
        ~~~~~~~~~+T_R n_f\biggl[ -\frac{17}{36} + \frac{2}{3}\ln(2D) \biggr]\biggr\} + \mathcal{O}(\as^2) \\ & \hspace{-1.2cm}\text{if}~~ f=g \notag \\
        1 + \frac{\as(\mu_R)}{\pi} \, C_F\biggl[ \frac{7}{4} -\frac{\pi^2}{4} -\frac{3}{2}\ln(2D) \\ ~~~~~~-\ln(D)\ln\biggl(\frac{Q^2}{4E^2_i} \biggr)-\ln^2(D)   \biggr] 
        + \mathcal{O}(\as^2) \\
        &\hspace{-1.2cm}\text{if}~~ f=q,\bar{q}
    \end{cases}
\end{align}
where $E_i$ is the jet energy in the partonic centre-of-mass frame.
The contribution $({\bf H S})_{cd}$ is given by
\begin{equation}
    ({\bf H S})_{cd} = \frac{\bra{\mathcal{M}_{cd}}{\bf S} \ket{\mathcal{M}_{cd}}}{|\mathcal{M}_{cd}^{(0)}|^2}
\end{equation}
where $\ket{\mathcal{M}_{cd}}$ is the UV renormalised virtual amplitude after the subtraction of infrared singularities\footnote{Our definition of the finite part of the one-loop amplitude corresponds to the conventions of Binoth Les Houches Accord \cite{Binoth:2010xt}.}, which admits a perturbative expansion in $\as(\mu_R)$.
The soft-parton factor ${\bf S}$ is an operator in colour space and can be expanded as
\begin{equation}
    {\bf S} = 1 + \frac{\as(\mu_R)}{\pi}{\bf S}^{(1)} + \mathcal{O}(\as^2)\, .
\end{equation}
The computation of the soft factor ${\bf S}^{(1)}$ can be carried out as follows.
We consider the emission of a soft gluon with momentum $k$ from the Born level process in Eq.~\eqref{eq:process}. We work in the partonic centre-of-mass frame and we parametrise the momentum $k$ as
\begin{equation}
  k=k_t\left(\cosh(\eta),\sin(\phi),\cos(\phi),\sinh(\eta)\right)\, .
  \end{equation}
We start from the squared eikonal current $\mathbf{J}^2=\mathbf{J}^2(k)$
and we reorganise it to explicitly subtract initial and final-state collinear singularities. We obtain  
\begin{widetext}
\begin{align}
  \label{eq:Jsub}
    \mathbf{J}^2_{\rm sub} &= \biggl( -\mathbf{T}_c\cdot \mathbf{T}_d\,\omega_{cd} 
    -\sum_{i}(\mathbf{T}_c\cdot \mathbf{T}_i\,\omega_{ci} + (c\leftrightarrow d) )
    -\sum_{i \ne j}\mathbf{T}_i\cdot \mathbf{T}_j\,\omega_{ij}    \biggr) \Theta(\rcut - k_{T,S}^{\rm ness}/Q) \notag \\
    &-\biggl(\mathbf{T}_c^2\,\omega^c_d + (c\leftrightarrow d) \biggr)\Theta(\rcut - k_t/Q) - \sum_i\,\mathbf{T}_i^2\,\omega_{C_iS}\Theta(\rcut - k_{T,C_iS}^{\rm ness}/Q)
\end{align}
\end{widetext}
where the sum runs over the labels of the final-state jets (i.e. $i,j=1,...,N$).
The eikonal kernels $\omega_{\alpha \beta}$, $\omega^{\alpha}_{\beta}$ are defined as
\begin{equation}
  \omega_{\alpha \beta} \equiv \frac{p_{\alpha} \cdot p_{\beta}}{(p_{\alpha}\cdot k)(p_{\beta}\cdot k)}
  ~~~~~~~\omega^{\alpha}_{\beta} \equiv \frac{p_{\alpha} \cdot p_{\beta}}{( p_{\alpha}\cdot k)\left((p_{\alpha} + p_{\beta})\cdot k\right)}
\end{equation}
and  $\omega_{C_iS}$ as
\begin{equation}
  \omega_{C_iS} \equiv  \frac{p_i \cdot (p_c + p_d)}{(p_i\cdot k)((p_c + p_d)\cdot k)} 
\end{equation}
where $\alpha$ and $\beta$ denote initial and/or final-state partons.
In the expression in Eq.~(\ref{eq:Jsub}) we have included phase space constraints that limit the integration to the region
$\ktness/Q<\rcut$. In particular $k_{T,S}^{\rm ness}$ is the soft limit of the resolution variable $\ktness$
\begin{equation}
    k_{T,S}^{\rm ness} = \min(1,\{\Delta R_{ik}\}/D)k_t
\end{equation}
while $k_{T,C_iS}^{\rm ness}$ is the soft limit of the $\ktness$ approximation used
in the final-state collinear limit and it can be expressed, in the partonic centre-of-mass frame, as
\begin{equation}
  (k_{T,C_iS}^{\rm ness})^2 =
  k_t^2\frac{\cosh(\eta)2(\cosh(\eta-y_i)-\cos(\phi-\phi_i))}{\cosh(y_i)D^2}\, .
\end{equation}
The integration of the initial state collinear contributions produce the customary collinear coefficient functions $C_{ca}$, while the integration of the final state collinear contribution produces the jet functions $J_i$. The leftover soft contributions can be obtained
by integrating the subtracted soft current $\mathbf{J}^2_{\rm sub}$ over the radiation phase space.
More precisely, the soft integrals in Eq.~\eqref{eq:Jsub} produce $1/\epsilon$ poles and logarithmic terms in $\rcut$.
In order to analytically extract them we define a new subtracted current as follows
\begin{equation}
    \mathbf{J}^2_{\rm sub} = \mathbf{J}^2_{\rm sing} + (\mathbf{J}^2_{\rm sub}-\mathbf{J}^2_{\rm sing}) \equiv \mathbf{J}^2_{\rm sing} + \mathbf{J}^2_{\rm fin}
\end{equation}
where $\mathbf{J}^2_{\rm sing}$ is still singular in the soft-wide-angle limit while 
\begin{equation}
  \left[\mathbf{J}^2_{\rm fin}\right]\equiv    8\pi^2\mu^{2\epsilon} \frac{1}{(2\pi)^{D-1}} \int d^Dk \delta_{+}(k^2)\mathbf{J}^2_{\rm fin}
\end{equation}
is finite in $D=4$ dimensions and can be computed numerically.\\
The soft-singular term $\mathbf{J}^2_{\rm sing} $ can be defined as
\begin{widetext}
\begin{align}
    \mathbf{J}^2_{\rm sing}  &= \sum_i\, \mathbf{T}_c \cdot \mathbf{T}_i \biggl( (\omega^c_d - \omega^c_i)\Theta(\rcut - k_t/Q)
    + (\omega_{C_iS} - \omega^i_c)\Theta(D\rcut - k_{i\perp}/Q) \biggr) + (c\leftrightarrow d) \notag \\
    &+ \sum_{i\ne j}\, \mathbf{T}_i \cdot \mathbf{T}_j (\omega_{C_iS} - \omega^i_j)\Theta(D\rcut - k_{i\perp}/Q)
\end{align}
\end{widetext}
where the sum runs over the labels of the final-state partons and $k_{i\perp}$ is the transverse momentum of $k$ with respect to the $i$-jet direction (in the partonic centre-of-mass frame):
\begin{widetext}
  \begin{equation}
  k^2_{i\perp}
    = k^2_t \frac{2(\cosh(\eta-y_i)-\cos(\phi-\phi_i)) (\cosh(\eta+y_i)+\cos(\phi-\phi_i))}{ \cosh(2y_i) +1} \,.
\end{equation}
\end{widetext}
The integration of $\mathbf{J}_{\rm sing}^2$ produces poles in $1/\epsilon$ and logarithmic terms in $\rcut$. The $1/\epsilon$ poles, together with the $1/\epsilon$ and $1/\epsilon^2$ poles coming from the initial- and final-state collinear integrals, cancel the corresponding poles in the virtual contribution. The logarithmic contributions in $\rcut$ produce the counterterm in Eq.~(4).
The finite remainder from the integration of $\mathbf{J}^2_{\rm sing} $ is 
\begin{widetext}
  \begin{align}
  \left[ \mathbf{J}^2_{\rm sing} \right]_{\rm fin}=
   -\frac{1}{2}\biggl\{
   &\sum_i\,\mathbf{T}_c \cdot \mathbf{T}_i\biggl[ 
     {\rm Li}_2\biggl(-\frac{(p_d\cdot p_i)}{(p_c\cdot p_d)} \biggr) +{\rm Li}_2\biggl(-\frac{(p_d\cdot p_i)(p_c\cdot p_d)}{((p_c+p_d)\cdot p_i)^2}\biggr)+2\ln(D)\ln\biggl( \frac{(p_c\cdot p_i)(p_c\cdot p_d)}{(p_i\cdot (p_c+p_d))^2}\biggr) \biggr] + (c\leftrightarrow d)\notag \\
   + &\sum_{i\ne j}\,\mathbf{T}_i \cdot \mathbf{T}_j\biggl[
   {\rm Li}_2\biggl( -\frac{((p_c+p_d)\cdot p_j)^2}{4(p_c\cdot p_d)(p_i\cdot p_j)}(1-\cos^2\theta_{ij}) \biggr)+ 2\ln(D) \ln\biggl( \frac{(p_i\cdot p_j)(p_c\cdot p_d)}{ p_i\cdot (p_c+p_d)p_j\cdot(p_c+p_d)} \biggr)
   \biggr]
   \biggr\}
\end{align}
\end{widetext}
where 
\begin{equation}
   \cos\theta_{ij} = 1-\frac{2(p_c\cdot p_d)(p_i\cdot p_j))}{p_i\cdot(p_c+p_d) p_j\cdot(p_c+p_d)} \,. 
\end{equation}
Finally, the soft factor ${\bf S}^{(1)}$ can be evaluated as
\begin{equation}
    {\bf S}^{(1)} =\left[ \mathbf{J}^2_{\rm sing} \right]_{\rm fin}+ \left[\mathbf{J}^2_{\rm fin}\right]\, .
\end{equation}

\bibliography{biblio}

\begin{thebibliography}{59}%
\makeatletter
\providecommand \@ifxundefined [1]{%
 \@ifx{#1\undefined}
}%
\providecommand \@ifnum [1]{%
 \ifnum #1\expandafter \@firstoftwo
 \else \expandafter \@secondoftwo
 \fi
}%
\providecommand \@ifx [1]{%
 \ifx #1\expandafter \@firstoftwo
 \else \expandafter \@secondoftwo
 \fi
}%
\providecommand \natexlab [1]{#1}%
\providecommand \enquote  [1]{``#1''}%
\providecommand \bibnamefont  [1]{#1}%
\providecommand \bibfnamefont [1]{#1}%
\providecommand \citenamefont [1]{#1}%
\providecommand \href@noop [0]{\@secondoftwo}%
\providecommand \href [0]{\begingroup \@sanitize@url \@href}%
\providecommand \@href[1]{\@@startlink{#1}\@@href}%
\providecommand \@@href[1]{\endgroup#1\@@endlink}%
\providecommand \@sanitize@url [0]{\catcode `\\12\catcode `\$12\catcode
  `\&12\catcode `\#12\catcode `\^12\catcode `\_12\catcode `\%12\relax}%
\providecommand \@@startlink[1]{}%
\providecommand \@@endlink[0]{}%
\providecommand \url  [0]{\begingroup\@sanitize@url \@url }%
\providecommand \@url [1]{\endgroup\@href {#1}{\urlprefix }}%
\providecommand \urlprefix  [0]{URL }%
\providecommand \Eprint [0]{\href }%
\providecommand \doibase [0]{https://doi.org/}%
\providecommand \selectlanguage [0]{\@gobble}%
\providecommand \bibinfo  [0]{\@secondoftwo}%
\providecommand \bibfield  [0]{\@secondoftwo}%
\providecommand \translation [1]{[#1]}%
\providecommand \BibitemOpen [0]{}%
\providecommand \bibitemStop [0]{}%
\providecommand \bibitemNoStop [0]{.\EOS\space}%
\providecommand \EOS [0]{\spacefactor3000\relax}%
\providecommand \BibitemShut  [1]{\csname bibitem#1\endcsname}%
\let\auto@bib@innerbib\@empty
\bibitem [{\citenamefont {Dasgupta}\ and\ \citenamefont
  {Salam}(2004)}]{Dasgupta:2003iq}%
  \BibitemOpen
  \bibfield  {author} {\bibinfo {author} {\bibfnamefont {M.}~\bibnamefont
  {Dasgupta}}\ and\ \bibinfo {author} {\bibfnamefont {G.~P.}\ \bibnamefont
  {Salam}},\ }\bibfield  {title} {\bibinfo {title} {{Event shapes in e+ e-
  annihilation and deep inelastic scattering}},\ }\href
  {https://doi.org/10.1088/0954-3899/30/5/R01} {\bibfield  {journal} {\bibinfo
  {journal} {J. Phys. G}\ }\textbf {\bibinfo {volume} {30}},\ \bibinfo {pages}
  {R143} (\bibinfo {year} {2004})},\ \Eprint
  {https://arxiv.org/abs/hep-ph/0312283} {arXiv:hep-ph/0312283} \BibitemShut
  {NoStop}%
\bibitem [{\citenamefont {Banfi}\ \emph {et~al.}(2004)\citenamefont {Banfi},
  \citenamefont {Salam},\ and\ \citenamefont {Zanderighi}}]{Banfi:2004nk}%
  \BibitemOpen
  \bibfield  {author} {\bibinfo {author} {\bibfnamefont {A.}~\bibnamefont
  {Banfi}}, \bibinfo {author} {\bibfnamefont {G.~P.}\ \bibnamefont {Salam}},\
  and\ \bibinfo {author} {\bibfnamefont {G.}~\bibnamefont {Zanderighi}},\
  }\bibfield  {title} {\bibinfo {title} {{Resummed event shapes at hadron -
  hadron colliders}},\ }\href {https://doi.org/10.1088/1126-6708/2004/08/062}
  {\bibfield  {journal} {\bibinfo  {journal} {JHEP}\ }\textbf {\bibinfo
  {volume} {08}},\ \bibinfo {pages} {062}},\ \Eprint
  {https://arxiv.org/abs/hep-ph/0407287} {arXiv:hep-ph/0407287} \BibitemShut
  {NoStop}%
\bibitem [{\citenamefont {Banfi}\ \emph {et~al.}(2010)\citenamefont {Banfi},
  \citenamefont {Salam},\ and\ \citenamefont {Zanderighi}}]{Banfi:2010xy}%
  \BibitemOpen
  \bibfield  {author} {\bibinfo {author} {\bibfnamefont {A.}~\bibnamefont
  {Banfi}}, \bibinfo {author} {\bibfnamefont {G.~P.}\ \bibnamefont {Salam}},\
  and\ \bibinfo {author} {\bibfnamefont {G.}~\bibnamefont {Zanderighi}},\
  }\bibfield  {title} {\bibinfo {title} {{Phenomenology of event shapes at
  hadron colliders}},\ }\href {https://doi.org/10.1007/JHEP06(2010)038}
  {\bibfield  {journal} {\bibinfo  {journal} {JHEP}\ }\textbf {\bibinfo
  {volume} {06}},\ \bibinfo {pages} {038}},\ \Eprint
  {https://arxiv.org/abs/1001.4082} {arXiv:1001.4082 [hep-ph]} \BibitemShut
  {NoStop}%
\bibitem [{\citenamefont {Stewart}\ \emph {et~al.}(2010)\citenamefont
  {Stewart}, \citenamefont {Tackmann},\ and\ \citenamefont
  {Waalewijn}}]{Stewart:2010tn}%
  \BibitemOpen
  \bibfield  {author} {\bibinfo {author} {\bibfnamefont {I.~W.}\ \bibnamefont
  {Stewart}}, \bibinfo {author} {\bibfnamefont {F.~J.}\ \bibnamefont
  {Tackmann}},\ and\ \bibinfo {author} {\bibfnamefont {W.~J.}\ \bibnamefont
  {Waalewijn}},\ }\bibfield  {title} {\bibinfo {title} {{N-Jettiness: An
  Inclusive Event Shape to Veto Jets}},\ }\href
  {https://doi.org/10.1103/PhysRevLett.105.092002} {\bibfield  {journal}
  {\bibinfo  {journal} {Phys. Rev. Lett.}\ }\textbf {\bibinfo {volume} {105}},\
  \bibinfo {pages} {092002} (\bibinfo {year} {2010})},\ \Eprint
  {https://arxiv.org/abs/1004.2489} {arXiv:1004.2489 [hep-ph]} \BibitemShut
  {NoStop}%
\bibitem [{\citenamefont {Catani}\ and\ \citenamefont
  {Grazzini}(2007)}]{Catani:2007vq}%
  \BibitemOpen
  \bibfield  {author} {\bibinfo {author} {\bibfnamefont {S.}~\bibnamefont
  {Catani}}\ and\ \bibinfo {author} {\bibfnamefont {M.}~\bibnamefont
  {Grazzini}},\ }\bibfield  {title} {\bibinfo {title} {{An NNLO subtraction
  formalism in hadron collisions and its application to Higgs boson production
  at the LHC}},\ }\href {https://doi.org/10.1103/PhysRevLett.98.222002}
  {\bibfield  {journal} {\bibinfo  {journal} {Phys. Rev. Lett.}\ }\textbf
  {\bibinfo {volume} {98}},\ \bibinfo {pages} {222002} (\bibinfo {year}
  {2007})},\ \Eprint {https://arxiv.org/abs/hep-ph/0703012}
  {arXiv:hep-ph/0703012} \BibitemShut {NoStop}%
\bibitem [{\citenamefont {Gaunt}\ \emph {et~al.}(2015)\citenamefont {Gaunt},
  \citenamefont {Stahlhofen}, \citenamefont {Tackmann},\ and\ \citenamefont
  {Walsh}}]{Gaunt:2015pea}%
  \BibitemOpen
  \bibfield  {author} {\bibinfo {author} {\bibfnamefont {J.}~\bibnamefont
  {Gaunt}}, \bibinfo {author} {\bibfnamefont {M.}~\bibnamefont {Stahlhofen}},
  \bibinfo {author} {\bibfnamefont {F.~J.}\ \bibnamefont {Tackmann}},\ and\
  \bibinfo {author} {\bibfnamefont {J.~R.}\ \bibnamefont {Walsh}},\ }\bibfield
  {title} {\bibinfo {title} {{N-jettiness Subtractions for NNLO QCD
  Calculations}},\ }\href {https://doi.org/10.1007/JHEP09(2015)058} {\bibfield
  {journal} {\bibinfo  {journal} {JHEP}\ }\textbf {\bibinfo {volume} {09}},\
  \bibinfo {pages} {058}},\ \Eprint {https://arxiv.org/abs/1505.04794}
  {arXiv:1505.04794 [hep-ph]} \BibitemShut {NoStop}%
\bibitem [{\citenamefont {Alioli}\ \emph {et~al.}(2014)\citenamefont {Alioli},
  \citenamefont {Bauer}, \citenamefont {Berggren}, \citenamefont {Tackmann},
  \citenamefont {Walsh},\ and\ \citenamefont {Zuberi}}]{Alioli:2013hqa}%
  \BibitemOpen
  \bibfield  {author} {\bibinfo {author} {\bibfnamefont {S.}~\bibnamefont
  {Alioli}}, \bibinfo {author} {\bibfnamefont {C.~W.}\ \bibnamefont {Bauer}},
  \bibinfo {author} {\bibfnamefont {C.}~\bibnamefont {Berggren}}, \bibinfo
  {author} {\bibfnamefont {F.~J.}\ \bibnamefont {Tackmann}}, \bibinfo {author}
  {\bibfnamefont {J.~R.}\ \bibnamefont {Walsh}},\ and\ \bibinfo {author}
  {\bibfnamefont {S.}~\bibnamefont {Zuberi}},\ }\bibfield  {title} {\bibinfo
  {title} {{Matching Fully Differential NNLO Calculations and Parton
  Showers}},\ }\href {https://doi.org/10.1007/JHEP06(2014)089} {\bibfield
  {journal} {\bibinfo  {journal} {JHEP}\ }\textbf {\bibinfo {volume} {06}},\
  \bibinfo {pages} {089}},\ \Eprint {https://arxiv.org/abs/1311.0286}
  {arXiv:1311.0286 [hep-ph]} \BibitemShut {NoStop}%
\bibitem [{\citenamefont {H\"oche}\ \emph {et~al.}(2014)\citenamefont
  {H\"oche}, \citenamefont {Li},\ and\ \citenamefont
  {Prestel}}]{Hoche:2014dla}%
  \BibitemOpen
  \bibfield  {author} {\bibinfo {author} {\bibfnamefont {S.}~\bibnamefont
  {H\"oche}}, \bibinfo {author} {\bibfnamefont {Y.}~\bibnamefont {Li}},\ and\
  \bibinfo {author} {\bibfnamefont {S.}~\bibnamefont {Prestel}},\ }\bibfield
  {title} {\bibinfo {title} {{Higgs-boson production through gluon fusion at
  NNLO QCD with parton showers}},\ }\href
  {https://doi.org/10.1103/PhysRevD.90.054011} {\bibfield  {journal} {\bibinfo
  {journal} {Phys. Rev. D}\ }\textbf {\bibinfo {volume} {90}},\ \bibinfo
  {pages} {054011} (\bibinfo {year} {2014})},\ \Eprint
  {https://arxiv.org/abs/1407.3773} {arXiv:1407.3773 [hep-ph]} \BibitemShut
  {NoStop}%
\bibitem [{\citenamefont {Monni}\ \emph {et~al.}(2020)\citenamefont {Monni},
  \citenamefont {Nason}, \citenamefont {Re}, \citenamefont {Wiesemann},\ and\
  \citenamefont {Zanderighi}}]{Monni:2019whf}%
  \BibitemOpen
  \bibfield  {author} {\bibinfo {author} {\bibfnamefont {P.~F.}\ \bibnamefont
  {Monni}}, \bibinfo {author} {\bibfnamefont {P.}~\bibnamefont {Nason}},
  \bibinfo {author} {\bibfnamefont {E.}~\bibnamefont {Re}}, \bibinfo {author}
  {\bibfnamefont {M.}~\bibnamefont {Wiesemann}},\ and\ \bibinfo {author}
  {\bibfnamefont {G.}~\bibnamefont {Zanderighi}},\ }\bibfield  {title}
  {\bibinfo {title} {{MiNNLO$_{PS}$: a new method to match NNLO QCD to parton
  showers}},\ }\href {https://doi.org/10.1007/JHEP05(2020)143} {\bibfield
  {journal} {\bibinfo  {journal} {JHEP}\ }\textbf {\bibinfo {volume} {05}},\
  \bibinfo {pages} {143}},\ \Eprint {https://arxiv.org/abs/1908.06987}
  {arXiv:1908.06987 [hep-ph]} \BibitemShut {NoStop}%
\bibitem [{\citenamefont {Mazzitelli}\ \emph {et~al.}(2021)\citenamefont
  {Mazzitelli}, \citenamefont {Monni}, \citenamefont {Nason}, \citenamefont
  {Re}, \citenamefont {Wiesemann},\ and\ \citenamefont
  {Zanderighi}}]{Mazzitelli:2020jio}%
  \BibitemOpen
  \bibfield  {author} {\bibinfo {author} {\bibfnamefont {J.}~\bibnamefont
  {Mazzitelli}}, \bibinfo {author} {\bibfnamefont {P.~F.}\ \bibnamefont
  {Monni}}, \bibinfo {author} {\bibfnamefont {P.}~\bibnamefont {Nason}},
  \bibinfo {author} {\bibfnamefont {E.}~\bibnamefont {Re}}, \bibinfo {author}
  {\bibfnamefont {M.}~\bibnamefont {Wiesemann}},\ and\ \bibinfo {author}
  {\bibfnamefont {G.}~\bibnamefont {Zanderighi}},\ }\bibfield  {title}
  {\bibinfo {title} {{Next-to-Next-to-Leading Order Event Generation for
  Top-Quark Pair Production}},\ }\href
  {https://doi.org/10.1103/PhysRevLett.127.062001} {\bibfield  {journal}
  {\bibinfo  {journal} {Phys. Rev. Lett.}\ }\textbf {\bibinfo {volume} {127}},\
  \bibinfo {pages} {062001} (\bibinfo {year} {2021})},\ \Eprint
  {https://arxiv.org/abs/2012.14267} {arXiv:2012.14267 [hep-ph]} \BibitemShut
  {NoStop}%
\bibitem [{\citenamefont {Alioli}\ \emph {et~al.}(2021)\citenamefont {Alioli},
  \citenamefont {Bauer}, \citenamefont {Broggio}, \citenamefont {Gavardi},
  \citenamefont {Kallweit}, \citenamefont {Lim}, \citenamefont {Nagar},
  \citenamefont {Napoletano},\ and\ \citenamefont {Rottoli}}]{Alioli:2021qbf}%
  \BibitemOpen
  \bibfield  {author} {\bibinfo {author} {\bibfnamefont {S.}~\bibnamefont
  {Alioli}}, \bibinfo {author} {\bibfnamefont {C.~W.}\ \bibnamefont {Bauer}},
  \bibinfo {author} {\bibfnamefont {A.}~\bibnamefont {Broggio}}, \bibinfo
  {author} {\bibfnamefont {A.}~\bibnamefont {Gavardi}}, \bibinfo {author}
  {\bibfnamefont {S.}~\bibnamefont {Kallweit}}, \bibinfo {author}
  {\bibfnamefont {M.~A.}\ \bibnamefont {Lim}}, \bibinfo {author} {\bibfnamefont
  {R.}~\bibnamefont {Nagar}}, \bibinfo {author} {\bibfnamefont
  {D.}~\bibnamefont {Napoletano}},\ and\ \bibinfo {author} {\bibfnamefont
  {L.}~\bibnamefont {Rottoli}},\ }\bibfield  {title} {\bibinfo {title}
  {{Matching NNLO predictions to parton showers using N3LL color-singlet
  transverse momentum resummation in GENEVA}},\ }\href
  {https://doi.org/10.1103/PhysRevD.104.094020} {\bibfield  {journal} {\bibinfo
   {journal} {Phys. Rev. D}\ }\textbf {\bibinfo {volume} {104}},\ \bibinfo
  {pages} {094020} (\bibinfo {year} {2021})},\ \Eprint
  {https://arxiv.org/abs/2102.08390} {arXiv:2102.08390 [hep-ph]} \BibitemShut
  {NoStop}%
\bibitem [{\citenamefont {Boughezal}\ \emph
  {et~al.}(2017{\natexlab{a}})\citenamefont {Boughezal}, \citenamefont
  {Campbell}, \citenamefont {Ellis}, \citenamefont {Focke}, \citenamefont
  {Giele}, \citenamefont {Liu}, \citenamefont {Petriello},\ and\ \citenamefont
  {Williams}}]{Boughezal:2016wmq}%
  \BibitemOpen
  \bibfield  {author} {\bibinfo {author} {\bibfnamefont {R.}~\bibnamefont
  {Boughezal}}, \bibinfo {author} {\bibfnamefont {J.~M.}\ \bibnamefont
  {Campbell}}, \bibinfo {author} {\bibfnamefont {R.~K.}\ \bibnamefont {Ellis}},
  \bibinfo {author} {\bibfnamefont {C.}~\bibnamefont {Focke}}, \bibinfo
  {author} {\bibfnamefont {W.}~\bibnamefont {Giele}}, \bibinfo {author}
  {\bibfnamefont {X.}~\bibnamefont {Liu}}, \bibinfo {author} {\bibfnamefont
  {F.}~\bibnamefont {Petriello}},\ and\ \bibinfo {author} {\bibfnamefont
  {C.}~\bibnamefont {Williams}},\ }\bibfield  {title} {\bibinfo {title} {{Color
  singlet production at NNLO in MCFM}},\ }\href
  {https://doi.org/10.1140/epjc/s10052-016-4558-y} {\bibfield  {journal}
  {\bibinfo  {journal} {Eur. Phys. J. C}\ }\textbf {\bibinfo {volume} {77}},\
  \bibinfo {pages} {7} (\bibinfo {year} {2017}{\natexlab{a}})},\ \Eprint
  {https://arxiv.org/abs/1605.08011} {arXiv:1605.08011 [hep-ph]} \BibitemShut
  {NoStop}%
\bibitem [{\citenamefont {Campbell}\ \emph {et~al.}(2016)\citenamefont
  {Campbell}, \citenamefont {Ellis}, \citenamefont {Li},\ and\ \citenamefont
  {Williams}}]{Campbell:2016yrh}%
  \BibitemOpen
  \bibfield  {author} {\bibinfo {author} {\bibfnamefont {J.~M.}\ \bibnamefont
  {Campbell}}, \bibinfo {author} {\bibfnamefont {R.~K.}\ \bibnamefont {Ellis}},
  \bibinfo {author} {\bibfnamefont {Y.}~\bibnamefont {Li}},\ and\ \bibinfo
  {author} {\bibfnamefont {C.}~\bibnamefont {Williams}},\ }\bibfield  {title}
  {\bibinfo {title} {{Predictions for diphoton production at the LHC through
  NNLO in QCD}},\ }\href {https://doi.org/10.1007/JHEP07(2016)148} {\bibfield
  {journal} {\bibinfo  {journal} {JHEP}\ }\textbf {\bibinfo {volume} {07}},\
  \bibinfo {pages} {148}},\ \Eprint {https://arxiv.org/abs/1603.02663}
  {arXiv:1603.02663 [hep-ph]} \BibitemShut {NoStop}%
\bibitem [{\citenamefont {Heinrich}\ \emph {et~al.}(2018)\citenamefont
  {Heinrich}, \citenamefont {Jahn}, \citenamefont {Jones}, \citenamefont
  {Kerner},\ and\ \citenamefont {Pires}}]{Heinrich:2017bvg}%
  \BibitemOpen
  \bibfield  {author} {\bibinfo {author} {\bibfnamefont {G.}~\bibnamefont
  {Heinrich}}, \bibinfo {author} {\bibfnamefont {S.}~\bibnamefont {Jahn}},
  \bibinfo {author} {\bibfnamefont {S.~P.}\ \bibnamefont {Jones}}, \bibinfo
  {author} {\bibfnamefont {M.}~\bibnamefont {Kerner}},\ and\ \bibinfo {author}
  {\bibfnamefont {J.}~\bibnamefont {Pires}},\ }\bibfield  {title} {\bibinfo
  {title} {{NNLO predictions for Z-boson pair production at the LHC}},\ }\href
  {https://doi.org/10.1007/JHEP03(2018)142} {\bibfield  {journal} {\bibinfo
  {journal} {JHEP}\ }\textbf {\bibinfo {volume} {03}},\ \bibinfo {pages}
  {142}},\ \Eprint {https://arxiv.org/abs/1710.06294} {arXiv:1710.06294
  [hep-ph]} \BibitemShut {NoStop}%
\bibitem [{\citenamefont {Campbell}\ \emph {et~al.}(2017)\citenamefont
  {Campbell}, \citenamefont {Neumann},\ and\ \citenamefont
  {Williams}}]{Campbell:2017aul}%
  \BibitemOpen
  \bibfield  {author} {\bibinfo {author} {\bibfnamefont {J.~M.}\ \bibnamefont
  {Campbell}}, \bibinfo {author} {\bibfnamefont {T.}~\bibnamefont {Neumann}},\
  and\ \bibinfo {author} {\bibfnamefont {C.}~\bibnamefont {Williams}},\
  }\bibfield  {title} {\bibinfo {title} {{$Z\gamma$ Production at NNLO
  Including Anomalous Couplings}},\ }\href
  {https://doi.org/10.1007/JHEP11(2017)150} {\bibfield  {journal} {\bibinfo
  {journal} {JHEP}\ }\textbf {\bibinfo {volume} {11}},\ \bibinfo {pages}
  {150}},\ \Eprint {https://arxiv.org/abs/1708.02925} {arXiv:1708.02925
  [hep-ph]} \BibitemShut {NoStop}%
\bibitem [{\citenamefont {Boughezal}\ \emph {et~al.}(2015)\citenamefont
  {Boughezal}, \citenamefont {Focke}, \citenamefont {Giele}, \citenamefont
  {Liu},\ and\ \citenamefont {Petriello}}]{Boughezal:2015aha}%
  \BibitemOpen
  \bibfield  {author} {\bibinfo {author} {\bibfnamefont {R.}~\bibnamefont
  {Boughezal}}, \bibinfo {author} {\bibfnamefont {C.}~\bibnamefont {Focke}},
  \bibinfo {author} {\bibfnamefont {W.}~\bibnamefont {Giele}}, \bibinfo
  {author} {\bibfnamefont {X.}~\bibnamefont {Liu}},\ and\ \bibinfo {author}
  {\bibfnamefont {F.}~\bibnamefont {Petriello}},\ }\bibfield  {title} {\bibinfo
  {title} {{Higgs boson production in association with a jet at NNLO using
  jettiness subtraction}},\ }\href
  {https://doi.org/10.1016/j.physletb.2015.06.055} {\bibfield  {journal}
  {\bibinfo  {journal} {Phys. Lett. B}\ }\textbf {\bibinfo {volume} {748}},\
  \bibinfo {pages} {5} (\bibinfo {year} {2015})},\ \Eprint
  {https://arxiv.org/abs/1505.03893} {arXiv:1505.03893 [hep-ph]} \BibitemShut
  {NoStop}%
\bibitem [{\citenamefont {Boughezal}\ \emph
  {et~al.}(2016{\natexlab{a}})\citenamefont {Boughezal}, \citenamefont
  {Campbell}, \citenamefont {Ellis}, \citenamefont {Focke}, \citenamefont
  {Giele}, \citenamefont {Liu},\ and\ \citenamefont
  {Petriello}}]{Boughezal:2015ded}%
  \BibitemOpen
  \bibfield  {author} {\bibinfo {author} {\bibfnamefont {R.}~\bibnamefont
  {Boughezal}}, \bibinfo {author} {\bibfnamefont {J.~M.}\ \bibnamefont
  {Campbell}}, \bibinfo {author} {\bibfnamefont {R.~K.}\ \bibnamefont {Ellis}},
  \bibinfo {author} {\bibfnamefont {C.}~\bibnamefont {Focke}}, \bibinfo
  {author} {\bibfnamefont {W.~T.}\ \bibnamefont {Giele}}, \bibinfo {author}
  {\bibfnamefont {X.}~\bibnamefont {Liu}},\ and\ \bibinfo {author}
  {\bibfnamefont {F.}~\bibnamefont {Petriello}},\ }\bibfield  {title} {\bibinfo
  {title} {{Z-boson production in association with a jet at
  next-to-next-to-leading order in perturbative QCD}},\ }\href
  {https://doi.org/10.1103/PhysRevLett.116.152001} {\bibfield  {journal}
  {\bibinfo  {journal} {Phys. Rev. Lett.}\ }\textbf {\bibinfo {volume} {116}},\
  \bibinfo {pages} {152001} (\bibinfo {year} {2016}{\natexlab{a}})},\ \Eprint
  {https://arxiv.org/abs/1512.01291} {arXiv:1512.01291 [hep-ph]} \BibitemShut
  {NoStop}%
\bibitem [{\citenamefont {Boughezal}\ \emph
  {et~al.}(2016{\natexlab{b}})\citenamefont {Boughezal}, \citenamefont {Liu},\
  and\ \citenamefont {Petriello}}]{Boughezal:2016dtm}%
  \BibitemOpen
  \bibfield  {author} {\bibinfo {author} {\bibfnamefont {R.}~\bibnamefont
  {Boughezal}}, \bibinfo {author} {\bibfnamefont {X.}~\bibnamefont {Liu}},\
  and\ \bibinfo {author} {\bibfnamefont {F.}~\bibnamefont {Petriello}},\
  }\bibfield  {title} {\bibinfo {title} {{W-boson plus jet differential
  distributions at NNLO in QCD}},\ }\href
  {https://doi.org/10.1103/PhysRevD.94.113009} {\bibfield  {journal} {\bibinfo
  {journal} {Phys. Rev. D}\ }\textbf {\bibinfo {volume} {94}},\ \bibinfo
  {pages} {113009} (\bibinfo {year} {2016}{\natexlab{b}})},\ \Eprint
  {https://arxiv.org/abs/1602.06965} {arXiv:1602.06965 [hep-ph]} \BibitemShut
  {NoStop}%
\bibitem [{\citenamefont {Moult}\ \emph {et~al.}(2017)\citenamefont {Moult},
  \citenamefont {Rothen}, \citenamefont {Stewart}, \citenamefont {Tackmann},\
  and\ \citenamefont {Zhu}}]{Moult:2016fqy}%
  \BibitemOpen
  \bibfield  {author} {\bibinfo {author} {\bibfnamefont {I.}~\bibnamefont
  {Moult}}, \bibinfo {author} {\bibfnamefont {L.}~\bibnamefont {Rothen}},
  \bibinfo {author} {\bibfnamefont {I.~W.}\ \bibnamefont {Stewart}}, \bibinfo
  {author} {\bibfnamefont {F.~J.}\ \bibnamefont {Tackmann}},\ and\ \bibinfo
  {author} {\bibfnamefont {H.~X.}\ \bibnamefont {Zhu}},\ }\bibfield  {title}
  {\bibinfo {title} {{Subleading Power Corrections for N-Jettiness
  Subtractions}},\ }\href {https://doi.org/10.1103/PhysRevD.95.074023}
  {\bibfield  {journal} {\bibinfo  {journal} {Phys. Rev. D}\ }\textbf {\bibinfo
  {volume} {95}},\ \bibinfo {pages} {074023} (\bibinfo {year} {2017})},\
  \Eprint {https://arxiv.org/abs/1612.00450} {arXiv:1612.00450 [hep-ph]}
  \BibitemShut {NoStop}%
\bibitem [{\citenamefont {Boughezal}\ \emph
  {et~al.}(2017{\natexlab{b}})\citenamefont {Boughezal}, \citenamefont {Liu},\
  and\ \citenamefont {Petriello}}]{Boughezal:2016zws}%
  \BibitemOpen
  \bibfield  {author} {\bibinfo {author} {\bibfnamefont {R.}~\bibnamefont
  {Boughezal}}, \bibinfo {author} {\bibfnamefont {X.}~\bibnamefont {Liu}},\
  and\ \bibinfo {author} {\bibfnamefont {F.}~\bibnamefont {Petriello}},\
  }\bibfield  {title} {\bibinfo {title} {{Power Corrections in the N-jettiness
  Subtraction Scheme}},\ }\href {https://doi.org/10.1007/JHEP03(2017)160}
  {\bibfield  {journal} {\bibinfo  {journal} {JHEP}\ }\textbf {\bibinfo
  {volume} {03}},\ \bibinfo {pages} {160}},\ \Eprint
  {https://arxiv.org/abs/1612.02911} {arXiv:1612.02911 [hep-ph]} \BibitemShut
  {NoStop}%
\bibitem [{\citenamefont {Boughezal}\ \emph {et~al.}(2018)\citenamefont
  {Boughezal}, \citenamefont {Isgr\`o},\ and\ \citenamefont
  {Petriello}}]{Boughezal:2018mvf}%
  \BibitemOpen
  \bibfield  {author} {\bibinfo {author} {\bibfnamefont {R.}~\bibnamefont
  {Boughezal}}, \bibinfo {author} {\bibfnamefont {A.}~\bibnamefont {Isgr\`o}},\
  and\ \bibinfo {author} {\bibfnamefont {F.}~\bibnamefont {Petriello}},\
  }\bibfield  {title} {\bibinfo {title} {{Next-to-leading-logarithmic power
  corrections for $N$-jettiness subtraction in color-singlet production}},\
  }\href {https://doi.org/10.1103/PhysRevD.97.076006} {\bibfield  {journal}
  {\bibinfo  {journal} {Phys. Rev. D}\ }\textbf {\bibinfo {volume} {97}},\
  \bibinfo {pages} {076006} (\bibinfo {year} {2018})},\ \Eprint
  {https://arxiv.org/abs/1802.00456} {arXiv:1802.00456 [hep-ph]} \BibitemShut
  {NoStop}%
\bibitem [{\citenamefont {Ebert}\ \emph {et~al.}(2018)\citenamefont {Ebert},
  \citenamefont {Moult}, \citenamefont {Stewart}, \citenamefont {Tackmann},
  \citenamefont {Vita},\ and\ \citenamefont {Zhu}}]{Ebert:2018lzn}%
  \BibitemOpen
  \bibfield  {author} {\bibinfo {author} {\bibfnamefont {M.~A.}\ \bibnamefont
  {Ebert}}, \bibinfo {author} {\bibfnamefont {I.}~\bibnamefont {Moult}},
  \bibinfo {author} {\bibfnamefont {I.~W.}\ \bibnamefont {Stewart}}, \bibinfo
  {author} {\bibfnamefont {F.~J.}\ \bibnamefont {Tackmann}}, \bibinfo {author}
  {\bibfnamefont {G.}~\bibnamefont {Vita}},\ and\ \bibinfo {author}
  {\bibfnamefont {H.~X.}\ \bibnamefont {Zhu}},\ }\bibfield  {title} {\bibinfo
  {title} {{Power Corrections for N-Jettiness Subtractions at ${\cal
  O}(\alpha_s)$}},\ }\href {https://doi.org/10.1007/JHEP12(2018)084} {\bibfield
   {journal} {\bibinfo  {journal} {JHEP}\ }\textbf {\bibinfo {volume} {12}},\
  \bibinfo {pages} {084}},\ \Eprint {https://arxiv.org/abs/1807.10764}
  {arXiv:1807.10764 [hep-ph]} \BibitemShut {NoStop}%
\bibitem [{\citenamefont {Campbell}\ \emph {et~al.}(2019)\citenamefont
  {Campbell}, \citenamefont {Ellis},\ and\ \citenamefont
  {Seth}}]{Campbell:2019gmd}%
  \BibitemOpen
  \bibfield  {author} {\bibinfo {author} {\bibfnamefont {J.~M.}\ \bibnamefont
  {Campbell}}, \bibinfo {author} {\bibfnamefont {R.~K.}\ \bibnamefont
  {Ellis}},\ and\ \bibinfo {author} {\bibfnamefont {S.}~\bibnamefont {Seth}},\
  }\bibfield  {title} {\bibinfo {title} {{H + 1 jet production revisited}},\
  }\href {https://doi.org/10.1007/JHEP10(2019)136} {\bibfield  {journal}
  {\bibinfo  {journal} {JHEP}\ }\textbf {\bibinfo {volume} {10}},\ \bibinfo
  {pages} {136}},\ \Eprint {https://arxiv.org/abs/1906.01020} {arXiv:1906.01020
  [hep-ph]} \BibitemShut {NoStop}%
\bibitem [{\citenamefont {Grazzini}\ \emph {et~al.}(2016)\citenamefont
  {Grazzini}, \citenamefont {Kallweit}, \citenamefont {Pozzorini},
  \citenamefont {Rathlev},\ and\ \citenamefont {Wiesemann}}]{Grazzini:2016ctr}%
  \BibitemOpen
  \bibfield  {author} {\bibinfo {author} {\bibfnamefont {M.}~\bibnamefont
  {Grazzini}}, \bibinfo {author} {\bibfnamefont {S.}~\bibnamefont {Kallweit}},
  \bibinfo {author} {\bibfnamefont {S.}~\bibnamefont {Pozzorini}}, \bibinfo
  {author} {\bibfnamefont {D.}~\bibnamefont {Rathlev}},\ and\ \bibinfo {author}
  {\bibfnamefont {M.}~\bibnamefont {Wiesemann}},\ }\bibfield  {title} {\bibinfo
  {title} {{$W^{+}W^{-}$ production at the LHC: fiducial cross sections and
  distributions in NNLO QCD}},\ }\href
  {https://doi.org/10.1007/JHEP08(2016)140} {\bibfield  {journal} {\bibinfo
  {journal} {JHEP}\ }\textbf {\bibinfo {volume} {08}},\ \bibinfo {pages}
  {140}},\ \Eprint {https://arxiv.org/abs/1605.02716} {arXiv:1605.02716
  [hep-ph]} \BibitemShut {NoStop}%
\bibitem [{\citenamefont {Ebert}\ \emph {et~al.}(2019)\citenamefont {Ebert},
  \citenamefont {Moult}, \citenamefont {Stewart}, \citenamefont {Tackmann},
  \citenamefont {Vita},\ and\ \citenamefont {Zhu}}]{Ebert:2018gsn}%
  \BibitemOpen
  \bibfield  {author} {\bibinfo {author} {\bibfnamefont {M.~A.}\ \bibnamefont
  {Ebert}}, \bibinfo {author} {\bibfnamefont {I.}~\bibnamefont {Moult}},
  \bibinfo {author} {\bibfnamefont {I.~W.}\ \bibnamefont {Stewart}}, \bibinfo
  {author} {\bibfnamefont {F.~J.}\ \bibnamefont {Tackmann}}, \bibinfo {author}
  {\bibfnamefont {G.}~\bibnamefont {Vita}},\ and\ \bibinfo {author}
  {\bibfnamefont {H.~X.}\ \bibnamefont {Zhu}},\ }\bibfield  {title} {\bibinfo
  {title} {{Subleading power rapidity divergences and power corrections for
  q$_{T}$}},\ }\href {https://doi.org/10.1007/JHEP04(2019)123} {\bibfield
  {journal} {\bibinfo  {journal} {JHEP}\ }\textbf {\bibinfo {volume} {04}},\
  \bibinfo {pages} {123}},\ \Eprint {https://arxiv.org/abs/1812.08189}
  {arXiv:1812.08189 [hep-ph]} \BibitemShut {NoStop}%
\bibitem [{\citenamefont {Cieri}\ \emph {et~al.}(2019)\citenamefont {Cieri},
  \citenamefont {Oleari},\ and\ \citenamefont {Rocco}}]{Cieri:2019tfv}%
  \BibitemOpen
  \bibfield  {author} {\bibinfo {author} {\bibfnamefont {L.}~\bibnamefont
  {Cieri}}, \bibinfo {author} {\bibfnamefont {C.}~\bibnamefont {Oleari}},\ and\
  \bibinfo {author} {\bibfnamefont {M.}~\bibnamefont {Rocco}},\ }\bibfield
  {title} {\bibinfo {title} {{Higher-order power corrections in a
  transverse-momentum cut for colour-singlet production at NLO}},\ }\href
  {https://doi.org/10.1140/epjc/s10052-019-7361-8} {\bibfield  {journal}
  {\bibinfo  {journal} {Eur. Phys. J. C}\ }\textbf {\bibinfo {volume} {79}},\
  \bibinfo {pages} {852} (\bibinfo {year} {2019})},\ \Eprint
  {https://arxiv.org/abs/1906.09044} {arXiv:1906.09044 [hep-ph]} \BibitemShut
  {NoStop}%
\bibitem [{\citenamefont {Grazzini}\ \emph {et~al.}(2018)\citenamefont
  {Grazzini}, \citenamefont {Kallweit},\ and\ \citenamefont
  {Wiesemann}}]{Grazzini:2017mhc}%
  \BibitemOpen
  \bibfield  {author} {\bibinfo {author} {\bibfnamefont {M.}~\bibnamefont
  {Grazzini}}, \bibinfo {author} {\bibfnamefont {S.}~\bibnamefont {Kallweit}},\
  and\ \bibinfo {author} {\bibfnamefont {M.}~\bibnamefont {Wiesemann}},\
  }\bibfield  {title} {\bibinfo {title} {{Fully differential NNLO computations
  with MATRIX}},\ }\href {https://doi.org/10.1140/epjc/s10052-018-5771-7}
  {\bibfield  {journal} {\bibinfo  {journal} {Eur. Phys. J. C}\ }\textbf
  {\bibinfo {volume} {78}},\ \bibinfo {pages} {537} (\bibinfo {year} {2018})},\
  \Eprint {https://arxiv.org/abs/1711.06631} {arXiv:1711.06631 [hep-ph]}
  \BibitemShut {NoStop}%
\bibitem [{\citenamefont {Buonocore}\ \emph {et~al.}(2020)\citenamefont
  {Buonocore}, \citenamefont {Grazzini},\ and\ \citenamefont
  {Tramontano}}]{Buonocore:2019puv}%
  \BibitemOpen
  \bibfield  {author} {\bibinfo {author} {\bibfnamefont {L.}~\bibnamefont
  {Buonocore}}, \bibinfo {author} {\bibfnamefont {M.}~\bibnamefont
  {Grazzini}},\ and\ \bibinfo {author} {\bibfnamefont {F.}~\bibnamefont
  {Tramontano}},\ }\bibfield  {title} {\bibinfo {title} {{The $q_T$ subtraction
  method: electroweak corrections and power suppressed contributions}},\ }\href
  {https://doi.org/10.1140/epjc/s10052-020-7815-z} {\bibfield  {journal}
  {\bibinfo  {journal} {Eur. Phys. J. C}\ }\textbf {\bibinfo {volume} {80}},\
  \bibinfo {pages} {254} (\bibinfo {year} {2020})},\ \Eprint
  {https://arxiv.org/abs/1911.10166} {arXiv:1911.10166 [hep-ph]} \BibitemShut
  {NoStop}%
\bibitem [{\citenamefont {Ebert}\ and\ \citenamefont
  {Tackmann}(2020)}]{Ebert:2019zkb}%
  \BibitemOpen
  \bibfield  {author} {\bibinfo {author} {\bibfnamefont {M.~A.}\ \bibnamefont
  {Ebert}}\ and\ \bibinfo {author} {\bibfnamefont {F.~J.}\ \bibnamefont
  {Tackmann}},\ }\bibfield  {title} {\bibinfo {title} {{Impact of isolation and
  fiducial cuts on q$_{T}$ and N-jettiness subtractions}},\ }\href
  {https://doi.org/10.1007/JHEP03(2020)158} {\bibfield  {journal} {\bibinfo
  {journal} {JHEP}\ }\textbf {\bibinfo {volume} {03}},\ \bibinfo {pages}
  {158}},\ \Eprint {https://arxiv.org/abs/1911.08486} {arXiv:1911.08486
  [hep-ph]} \BibitemShut {NoStop}%
\bibitem [{\citenamefont {Catani}\ \emph
  {et~al.}(2019{\natexlab{a}})\citenamefont {Catani}, \citenamefont {Devoto},
  \citenamefont {Grazzini}, \citenamefont {Kallweit}, \citenamefont
  {Mazzitelli},\ and\ \citenamefont {Sargsyan}}]{Catani:2019iny}%
  \BibitemOpen
  \bibfield  {author} {\bibinfo {author} {\bibfnamefont {S.}~\bibnamefont
  {Catani}}, \bibinfo {author} {\bibfnamefont {S.}~\bibnamefont {Devoto}},
  \bibinfo {author} {\bibfnamefont {M.}~\bibnamefont {Grazzini}}, \bibinfo
  {author} {\bibfnamefont {S.}~\bibnamefont {Kallweit}}, \bibinfo {author}
  {\bibfnamefont {J.}~\bibnamefont {Mazzitelli}},\ and\ \bibinfo {author}
  {\bibfnamefont {H.}~\bibnamefont {Sargsyan}},\ }\bibfield  {title} {\bibinfo
  {title} {{Top-quark pair hadroproduction at next-to-next-to-leading order in
  QCD}},\ }\href {https://doi.org/10.1103/PhysRevD.99.051501} {\bibfield
  {journal} {\bibinfo  {journal} {Phys. Rev. D}\ }\textbf {\bibinfo {volume}
  {99}},\ \bibinfo {pages} {051501} (\bibinfo {year} {2019}{\natexlab{a}})},\
  \Eprint {https://arxiv.org/abs/1901.04005} {arXiv:1901.04005 [hep-ph]}
  \BibitemShut {NoStop}%
\bibitem [{\citenamefont {Gaunt}(2014)}]{Gaunt:2014ska}%
  \BibitemOpen
  \bibfield  {author} {\bibinfo {author} {\bibfnamefont {J.~R.}\ \bibnamefont
  {Gaunt}},\ }\bibfield  {title} {\bibinfo {title} {{Glauber Gluons and
  Multiple Parton Interactions}},\ }\href
  {https://doi.org/10.1007/JHEP07(2014)110} {\bibfield  {journal} {\bibinfo
  {journal} {JHEP}\ }\textbf {\bibinfo {volume} {07}},\ \bibinfo {pages}
  {110}},\ \Eprint {https://arxiv.org/abs/1405.2080} {arXiv:1405.2080 [hep-ph]}
  \BibitemShut {NoStop}%
\bibitem [{\citenamefont {Catani}\ \emph {et~al.}(1993)\citenamefont {Catani},
  \citenamefont {Dokshitzer}, \citenamefont {Seymour},\ and\ \citenamefont
  {Webber}}]{Catani:1993hr}%
  \BibitemOpen
  \bibfield  {author} {\bibinfo {author} {\bibfnamefont {S.}~\bibnamefont
  {Catani}}, \bibinfo {author} {\bibfnamefont {Y.~L.}\ \bibnamefont
  {Dokshitzer}}, \bibinfo {author} {\bibfnamefont {M.~H.}\ \bibnamefont
  {Seymour}},\ and\ \bibinfo {author} {\bibfnamefont {B.~R.}\ \bibnamefont
  {Webber}},\ }\bibfield  {title} {\bibinfo {title} {{Longitudinally invariant
  $K_t$ clustering algorithms for hadron hadron collisions}},\ }\href
  {https://doi.org/10.1016/0550-3213(93)90166-M} {\bibfield  {journal}
  {\bibinfo  {journal} {Nucl. Phys. B}\ }\textbf {\bibinfo {volume} {406}},\
  \bibinfo {pages} {187} (\bibinfo {year} {1993})}\BibitemShut {NoStop}%
\bibitem [{\citenamefont {Ellis}\ and\ \citenamefont
  {Soper}(1993)}]{Ellis:1993tq}%
  \BibitemOpen
  \bibfield  {author} {\bibinfo {author} {\bibfnamefont {S.~D.}\ \bibnamefont
  {Ellis}}\ and\ \bibinfo {author} {\bibfnamefont {D.~E.}\ \bibnamefont
  {Soper}},\ }\bibfield  {title} {\bibinfo {title} {{Successive combination jet
  algorithm for hadron collisions}},\ }\href
  {https://doi.org/10.1103/PhysRevD.48.3160} {\bibfield  {journal} {\bibinfo
  {journal} {Phys. Rev. D}\ }\textbf {\bibinfo {volume} {48}},\ \bibinfo
  {pages} {3160} (\bibinfo {year} {1993})},\ \Eprint
  {https://arxiv.org/abs/hep-ph/9305266} {arXiv:hep-ph/9305266} \BibitemShut
  {NoStop}%
\bibitem [{\citenamefont {Cacciari}\ \emph {et~al.}(2012)\citenamefont
  {Cacciari}, \citenamefont {Salam},\ and\ \citenamefont
  {Soyez}}]{Cacciari:2011ma}%
  \BibitemOpen
  \bibfield  {author} {\bibinfo {author} {\bibfnamefont {M.}~\bibnamefont
  {Cacciari}}, \bibinfo {author} {\bibfnamefont {G.~P.}\ \bibnamefont
  {Salam}},\ and\ \bibinfo {author} {\bibfnamefont {G.}~\bibnamefont {Soyez}},\
  }\bibfield  {title} {\bibinfo {title} {{FastJet User Manual}},\ }\href
  {https://doi.org/10.1140/epjc/s10052-012-1896-2} {\bibfield  {journal}
  {\bibinfo  {journal} {Eur. Phys. J. C}\ }\textbf {\bibinfo {volume} {72}},\
  \bibinfo {pages} {1896} (\bibinfo {year} {2012})},\ \Eprint
  {https://arxiv.org/abs/1111.6097} {arXiv:1111.6097 [hep-ph]} \BibitemShut
  {NoStop}%
\bibitem [{\citenamefont {Catani}\ \emph {et~al.}(2014)\citenamefont {Catani},
  \citenamefont {Grazzini},\ and\ \citenamefont {Torre}}]{Catani:2014qha}%
  \BibitemOpen
  \bibfield  {author} {\bibinfo {author} {\bibfnamefont {S.}~\bibnamefont
  {Catani}}, \bibinfo {author} {\bibfnamefont {M.}~\bibnamefont {Grazzini}},\
  and\ \bibinfo {author} {\bibfnamefont {A.}~\bibnamefont {Torre}},\ }\bibfield
   {title} {\bibinfo {title} {{Transverse-momentum resummation for heavy-quark
  hadroproduction}},\ }\href {https://doi.org/10.1016/j.nuclphysb.2014.11.019}
  {\bibfield  {journal} {\bibinfo  {journal} {Nucl. Phys. B}\ }\textbf
  {\bibinfo {volume} {890}},\ \bibinfo {pages} {518} (\bibinfo {year}
  {2014})},\ \Eprint {https://arxiv.org/abs/1408.4564} {arXiv:1408.4564
  [hep-ph]} \BibitemShut {NoStop}%
\bibitem [{\citenamefont {Buonocore}\ \emph {et~al.}(2022)\citenamefont
  {Buonocore}, \citenamefont {Grazzini}, \citenamefont {Haag},\ and\
  \citenamefont {Rottoli}}]{Buonocore:2021akg}%
  \BibitemOpen
  \bibfield  {author} {\bibinfo {author} {\bibfnamefont {L.}~\bibnamefont
  {Buonocore}}, \bibinfo {author} {\bibfnamefont {M.}~\bibnamefont {Grazzini}},
  \bibinfo {author} {\bibfnamefont {J.}~\bibnamefont {Haag}},\ and\ \bibinfo
  {author} {\bibfnamefont {L.}~\bibnamefont {Rottoli}},\ }\bibfield  {title}
  {\bibinfo {title} {{Transverse-momentum resummation for boson plus jet
  production at hadron colliders}},\ }\href
  {https://doi.org/10.1140/epjc/s10052-021-09962-4} {\bibfield  {journal}
  {\bibinfo  {journal} {Eur. Phys. J. C}\ }\textbf {\bibinfo {volume} {82}},\
  \bibinfo {pages} {27} (\bibinfo {year} {2022})},\ \Eprint
  {https://arxiv.org/abs/2110.06913} {arXiv:2110.06913 [hep-ph]} \BibitemShut
  {NoStop}%
\bibitem [{\citenamefont {Altarelli}\ and\ \citenamefont
  {Parisi}(1977)}]{Altarelli:1977zs}%
  \BibitemOpen
  \bibfield  {author} {\bibinfo {author} {\bibfnamefont {G.}~\bibnamefont
  {Altarelli}}\ and\ \bibinfo {author} {\bibfnamefont {G.}~\bibnamefont
  {Parisi}},\ }\bibfield  {title} {\bibinfo {title} {{Asymptotic Freedom in
  Parton Language}},\ }\href {https://doi.org/10.1016/0550-3213(77)90384-4}
  {\bibfield  {journal} {\bibinfo  {journal} {Nucl. Phys. B}\ }\textbf
  {\bibinfo {volume} {126}},\ \bibinfo {pages} {298} (\bibinfo {year}
  {1977})}\BibitemShut {NoStop}%
\bibitem [{\citenamefont {Dokshitzer}(1977)}]{Dokshitzer:1977sg}%
  \BibitemOpen
  \bibfield  {author} {\bibinfo {author} {\bibfnamefont {Y.~L.}\ \bibnamefont
  {Dokshitzer}},\ }\bibfield  {title} {\bibinfo {title} {{Calculation of the
  Structure Functions for Deep Inelastic Scattering and e+ e- Annihilation by
  Perturbation Theory in Quantum Chromodynamics.}},\ }\href@noop {} {\bibfield
  {journal} {\bibinfo  {journal} {Sov. Phys. JETP}\ }\textbf {\bibinfo {volume}
  {46}},\ \bibinfo {pages} {641} (\bibinfo {year} {1977})}\BibitemShut
  {NoStop}%
\bibitem [{\citenamefont {Gribov}\ and\ \citenamefont
  {Lipatov}(1972)}]{Gribov:1972ri}%
  \BibitemOpen
  \bibfield  {author} {\bibinfo {author} {\bibfnamefont {V.~N.}\ \bibnamefont
  {Gribov}}\ and\ \bibinfo {author} {\bibfnamefont {L.~N.}\ \bibnamefont
  {Lipatov}},\ }\bibfield  {title} {\bibinfo {title} {{Deep inelastic e p
  scattering in perturbation theory}},\ }\href@noop {} {\bibfield  {journal}
  {\bibinfo  {journal} {Sov. J. Nucl. Phys.}\ }\textbf {\bibinfo {volume}
  {15}},\ \bibinfo {pages} {438} (\bibinfo {year} {1972})}\BibitemShut
  {NoStop}%
\bibitem [{\citenamefont {Campbell}\ and\ \citenamefont
  {Neumann}(2019)}]{Campbell:2019dru}%
  \BibitemOpen
  \bibfield  {author} {\bibinfo {author} {\bibfnamefont {J.}~\bibnamefont
  {Campbell}}\ and\ \bibinfo {author} {\bibfnamefont {T.}~\bibnamefont
  {Neumann}},\ }\bibfield  {title} {\bibinfo {title} {{Precision Phenomenology
  with MCFM}},\ }\href {https://doi.org/10.1007/JHEP12(2019)034} {\bibfield
  {journal} {\bibinfo  {journal} {JHEP}\ }\textbf {\bibinfo {volume} {12}},\
  \bibinfo {pages} {034}},\ \Eprint {https://arxiv.org/abs/1909.09117}
  {arXiv:1909.09117 [hep-ph]} \BibitemShut {NoStop}%
\bibitem [{\citenamefont {Cascioli}\ \emph {et~al.}(2012)\citenamefont
  {Cascioli}, \citenamefont {Maierh{\"o}fer},\ and\ \citenamefont
  {Pozzorini}}]{Cascioli:2011va}%
  \BibitemOpen
  \bibfield  {author} {\bibinfo {author} {\bibfnamefont {F.}~\bibnamefont
  {Cascioli}}, \bibinfo {author} {\bibfnamefont {P.}~\bibnamefont
  {Maierh{\"o}fer}},\ and\ \bibinfo {author} {\bibfnamefont {S.}~\bibnamefont
  {Pozzorini}},\ }\bibfield  {title} {\bibinfo {title} {{Scattering Amplitudes
  with Open Loops}},\ }\href {https://doi.org/10.1103/PhysRevLett.108.111601}
  {\bibfield  {journal} {\bibinfo  {journal} {Phys. Rev. Lett.}\ }\textbf
  {\bibinfo {volume} {108}},\ \bibinfo {pages} {111601} (\bibinfo {year}
  {2012})},\ \Eprint {https://arxiv.org/abs/1111.5206} {arXiv:1111.5206
  [hep-ph]} \BibitemShut {NoStop}%
\bibitem [{\citenamefont {Buccioni}\ \emph {et~al.}(2018)\citenamefont
  {Buccioni}, \citenamefont {Pozzorini},\ and\ \citenamefont
  {Zoller}}]{Buccioni:2017yxi}%
  \BibitemOpen
  \bibfield  {author} {\bibinfo {author} {\bibfnamefont {F.}~\bibnamefont
  {Buccioni}}, \bibinfo {author} {\bibfnamefont {S.}~\bibnamefont
  {Pozzorini}},\ and\ \bibinfo {author} {\bibfnamefont {M.}~\bibnamefont
  {Zoller}},\ }\bibfield  {title} {\bibinfo {title} {{On-the-fly reduction of
  open loops}},\ }\href {https://doi.org/10.1140/epjc/s10052-018-5562-1}
  {\bibfield  {journal} {\bibinfo  {journal} {Eur. Phys. J. C}\ }\textbf
  {\bibinfo {volume} {78}},\ \bibinfo {pages} {70} (\bibinfo {year} {2018})},\
  \Eprint {https://arxiv.org/abs/1710.11452} {arXiv:1710.11452 [hep-ph]}
  \BibitemShut {NoStop}%
\bibitem [{\citenamefont {Buccioni}\ \emph {et~al.}(2019)\citenamefont
  {Buccioni}, \citenamefont {Lang}, \citenamefont {Lindert}, \citenamefont
  {Maierh{\"o}fer}, \citenamefont {Pozzorini}, \citenamefont {Zhang},\ and\
  \citenamefont {Zoller}}]{Buccioni:2019sur}%
  \BibitemOpen
  \bibfield  {author} {\bibinfo {author} {\bibfnamefont {F.}~\bibnamefont
  {Buccioni}}, \bibinfo {author} {\bibfnamefont {J.-N.}\ \bibnamefont {Lang}},
  \bibinfo {author} {\bibfnamefont {J.~M.}\ \bibnamefont {Lindert}}, \bibinfo
  {author} {\bibfnamefont {P.}~\bibnamefont {Maierh{\"o}fer}}, \bibinfo
  {author} {\bibfnamefont {S.}~\bibnamefont {Pozzorini}}, \bibinfo {author}
  {\bibfnamefont {H.}~\bibnamefont {Zhang}},\ and\ \bibinfo {author}
  {\bibfnamefont {M.~F.}\ \bibnamefont {Zoller}},\ }\bibfield  {title}
  {\bibinfo {title} {{OpenLoops 2}},\ }\href
  {https://doi.org/10.1140/epjc/s10052-019-7306-2} {\bibfield  {journal}
  {\bibinfo  {journal} {Eur. Phys. J. C}\ }\textbf {\bibinfo {volume} {79}},\
  \bibinfo {pages} {866} (\bibinfo {year} {2019})},\ \Eprint
  {https://arxiv.org/abs/1907.13071} {arXiv:1907.13071 [hep-ph]} \BibitemShut
  {NoStop}%
\bibitem [{\citenamefont {Ball}\ \emph {et~al.}(2017)\citenamefont {Ball} \emph
  {et~al.}}]{NNPDF:2017mvq}%
  \BibitemOpen
  \bibfield  {author} {\bibinfo {author} {\bibfnamefont {R.~D.}\ \bibnamefont
  {Ball}} \emph {et~al.} (\bibinfo {collaboration} {NNPDF}),\ }\bibfield
  {title} {\bibinfo {title} {{Parton distributions from high-precision collider
  data}},\ }\href {https://doi.org/10.1140/epjc/s10052-017-5199-5} {\bibfield
  {journal} {\bibinfo  {journal} {Eur. Phys. J. C}\ }\textbf {\bibinfo {volume}
  {77}},\ \bibinfo {pages} {663} (\bibinfo {year} {2017})},\ \Eprint
  {https://arxiv.org/abs/1706.00428} {arXiv:1706.00428 [hep-ph]} \BibitemShut
  {NoStop}%
\bibitem [{\citenamefont {Buckley}\ \emph {et~al.}(2015)\citenamefont
  {Buckley}, \citenamefont {Ferrando}, \citenamefont {Lloyd}, \citenamefont
  {Nordstr\"om}, \citenamefont {Page}, \citenamefont {R\"ufenacht},
  \citenamefont {Sch\"onherr},\ and\ \citenamefont {Watt}}]{Buckley:2014ana}%
  \BibitemOpen
  \bibfield  {author} {\bibinfo {author} {\bibfnamefont {A.}~\bibnamefont
  {Buckley}}, \bibinfo {author} {\bibfnamefont {J.}~\bibnamefont {Ferrando}},
  \bibinfo {author} {\bibfnamefont {S.}~\bibnamefont {Lloyd}}, \bibinfo
  {author} {\bibfnamefont {K.}~\bibnamefont {Nordstr\"om}}, \bibinfo {author}
  {\bibfnamefont {B.}~\bibnamefont {Page}}, \bibinfo {author} {\bibfnamefont
  {M.}~\bibnamefont {R\"ufenacht}}, \bibinfo {author} {\bibfnamefont
  {M.}~\bibnamefont {Sch\"onherr}},\ and\ \bibinfo {author} {\bibfnamefont
  {G.}~\bibnamefont {Watt}},\ }\bibfield  {title} {\bibinfo {title} {{LHAPDF6:
  parton density access in the LHC precision era}},\ }\href
  {https://doi.org/10.1140/epjc/s10052-015-3318-8} {\bibfield  {journal}
  {\bibinfo  {journal} {Eur. Phys. J. C}\ }\textbf {\bibinfo {volume} {75}},\
  \bibinfo {pages} {132} (\bibinfo {year} {2015})},\ \Eprint
  {https://arxiv.org/abs/1412.7420} {arXiv:1412.7420 [hep-ph]} \BibitemShut
  {NoStop}%
\bibitem [{\citenamefont {Cacciari}\ \emph {et~al.}(2008)\citenamefont
  {Cacciari}, \citenamefont {Salam},\ and\ \citenamefont
  {Soyez}}]{Cacciari:2008gp}%
  \BibitemOpen
  \bibfield  {author} {\bibinfo {author} {\bibfnamefont {M.}~\bibnamefont
  {Cacciari}}, \bibinfo {author} {\bibfnamefont {G.~P.}\ \bibnamefont
  {Salam}},\ and\ \bibinfo {author} {\bibfnamefont {G.}~\bibnamefont {Soyez}},\
  }\bibfield  {title} {\bibinfo {title} {{The anti-$k_t$ jet clustering
  algorithm}},\ }\href {https://doi.org/10.1088/1126-6708/2008/04/063}
  {\bibfield  {journal} {\bibinfo  {journal} {JHEP}\ }\textbf {\bibinfo
  {volume} {04}},\ \bibinfo {pages} {063}},\ \Eprint
  {https://arxiv.org/abs/0802.1189} {arXiv:0802.1189 [hep-ph]} \BibitemShut
  {NoStop}%
\bibitem [{\citenamefont {Catani}\ and\ \citenamefont
  {Seymour}(1996)}]{Catani:1996jh}%
  \BibitemOpen
  \bibfield  {author} {\bibinfo {author} {\bibfnamefont {S.}~\bibnamefont
  {Catani}}\ and\ \bibinfo {author} {\bibfnamefont {M.~H.}\ \bibnamefont
  {Seymour}},\ }\bibfield  {title} {\bibinfo {title} {{The Dipole formalism for
  the calculation of QCD jet cross-sections at next-to-leading order}},\ }\href
  {https://doi.org/10.1016/0370-2693(96)00425-X} {\bibfield  {journal}
  {\bibinfo  {journal} {Phys. Lett. B}\ }\textbf {\bibinfo {volume} {378}},\
  \bibinfo {pages} {287} (\bibinfo {year} {1996})},\ \Eprint
  {https://arxiv.org/abs/hep-ph/9602277} {arXiv:hep-ph/9602277} \BibitemShut
  {NoStop}%
\bibitem [{\citenamefont {Catani}\ and\ \citenamefont
  {Seymour}(1997)}]{Catani:1996vz}%
  \BibitemOpen
  \bibfield  {author} {\bibinfo {author} {\bibfnamefont {S.}~\bibnamefont
  {Catani}}\ and\ \bibinfo {author} {\bibfnamefont {M.~H.}\ \bibnamefont
  {Seymour}},\ }\bibfield  {title} {\bibinfo {title} {{A General algorithm for
  calculating jet cross-sections in NLO QCD}},\ }\href
  {https://doi.org/10.1016/S0550-3213(96)00589-5} {\bibfield  {journal}
  {\bibinfo  {journal} {Nucl. Phys. B}\ }\textbf {\bibinfo {volume} {485}},\
  \bibinfo {pages} {291} (\bibinfo {year} {1997})},\ \bibinfo {note} {[Erratum:
  Nucl.Phys.B 510, 503--504 (1998)]},\ \Eprint
  {https://arxiv.org/abs/hep-ph/9605323} {arXiv:hep-ph/9605323} \BibitemShut
  {NoStop}%
\bibitem [{\citenamefont {Kallweit}\ \emph {et~al.}(2016)\citenamefont
  {Kallweit}, \citenamefont {Lindert}, \citenamefont {Maierhofer},
  \citenamefont {Pozzorini},\ and\ \citenamefont
  {Sch\"onherr}}]{Kallweit:2015dum}%
  \BibitemOpen
  \bibfield  {author} {\bibinfo {author} {\bibfnamefont {S.}~\bibnamefont
  {Kallweit}}, \bibinfo {author} {\bibfnamefont {J.~M.}\ \bibnamefont
  {Lindert}}, \bibinfo {author} {\bibfnamefont {P.}~\bibnamefont {Maierhofer}},
  \bibinfo {author} {\bibfnamefont {S.}~\bibnamefont {Pozzorini}},\ and\
  \bibinfo {author} {\bibfnamefont {M.}~\bibnamefont {Sch\"onherr}},\
  }\bibfield  {title} {\bibinfo {title} {{NLO QCD+EW predictions for V + jets
  including off-shell vector-boson decays and multijet merging}},\ }\href
  {https://doi.org/10.1007/JHEP04(2016)021} {\bibfield  {journal} {\bibinfo
  {journal} {JHEP}\ }\textbf {\bibinfo {volume} {04}},\ \bibinfo {pages}
  {021}},\ \Eprint {https://arxiv.org/abs/1511.08692} {arXiv:1511.08692
  [hep-ph]} \BibitemShut {NoStop}%
\bibitem [{\citenamefont {Nason}(2004)}]{Nason:2004rx}%
  \BibitemOpen
  \bibfield  {author} {\bibinfo {author} {\bibfnamefont {P.}~\bibnamefont
  {Nason}},\ }\bibfield  {title} {\bibinfo {title} {{A New method for combining
  NLO QCD with shower Monte Carlo algorithms}},\ }\href
  {https://doi.org/10.1088/1126-6708/2004/11/040} {\bibfield  {journal}
  {\bibinfo  {journal} {JHEP}\ }\textbf {\bibinfo {volume} {11}},\ \bibinfo
  {pages} {040}},\ \Eprint {https://arxiv.org/abs/hep-ph/0409146}
  {arXiv:hep-ph/0409146} \BibitemShut {NoStop}%
\bibitem [{\citenamefont {Frixione}\ \emph {et~al.}(2007)\citenamefont
  {Frixione}, \citenamefont {Nason},\ and\ \citenamefont
  {Oleari}}]{Frixione:2007vw}%
  \BibitemOpen
  \bibfield  {author} {\bibinfo {author} {\bibfnamefont {S.}~\bibnamefont
  {Frixione}}, \bibinfo {author} {\bibfnamefont {P.}~\bibnamefont {Nason}},\
  and\ \bibinfo {author} {\bibfnamefont {C.}~\bibnamefont {Oleari}},\
  }\bibfield  {title} {\bibinfo {title} {{Matching NLO QCD computations with
  Parton Shower simulations: the POWHEG method}},\ }\href
  {https://doi.org/10.1088/1126-6708/2007/11/070} {\bibfield  {journal}
  {\bibinfo  {journal} {JHEP}\ }\textbf {\bibinfo {volume} {11}},\ \bibinfo
  {pages} {070}},\ \Eprint {https://arxiv.org/abs/0709.2092} {arXiv:0709.2092
  [hep-ph]} \BibitemShut {NoStop}%
\bibitem [{\citenamefont {Alioli}\ \emph {et~al.}(2011)\citenamefont {Alioli},
  \citenamefont {Nason}, \citenamefont {Oleari},\ and\ \citenamefont
  {Re}}]{Alioli:2010qp}%
  \BibitemOpen
  \bibfield  {author} {\bibinfo {author} {\bibfnamefont {S.}~\bibnamefont
  {Alioli}}, \bibinfo {author} {\bibfnamefont {P.}~\bibnamefont {Nason}},
  \bibinfo {author} {\bibfnamefont {C.}~\bibnamefont {Oleari}},\ and\ \bibinfo
  {author} {\bibfnamefont {E.}~\bibnamefont {Re}},\ }\bibfield  {title}
  {\bibinfo {title} {{Vector boson plus one jet production in POWHEG}},\ }\href
  {https://doi.org/10.1007/JHEP01(2011)095} {\bibfield  {journal} {\bibinfo
  {journal} {JHEP}\ }\textbf {\bibinfo {volume} {01}},\ \bibinfo {pages}
  {095}},\ \Eprint {https://arxiv.org/abs/1009.5594} {arXiv:1009.5594 [hep-ph]}
  \BibitemShut {NoStop}%
\bibitem [{\citenamefont {Sj\"ostrand}\ \emph {et~al.}(2015)\citenamefont
  {Sj\"ostrand}, \citenamefont {Ask}, \citenamefont {Christiansen},
  \citenamefont {Corke}, \citenamefont {Desai}, \citenamefont {Ilten},
  \citenamefont {Mrenna}, \citenamefont {Prestel}, \citenamefont {Rasmussen},\
  and\ \citenamefont {Skands}}]{Sjostrand:2014zea}%
  \BibitemOpen
  \bibfield  {author} {\bibinfo {author} {\bibfnamefont {T.}~\bibnamefont
  {Sj\"ostrand}}, \bibinfo {author} {\bibfnamefont {S.}~\bibnamefont {Ask}},
  \bibinfo {author} {\bibfnamefont {J.~R.}\ \bibnamefont {Christiansen}},
  \bibinfo {author} {\bibfnamefont {R.}~\bibnamefont {Corke}}, \bibinfo
  {author} {\bibfnamefont {N.}~\bibnamefont {Desai}}, \bibinfo {author}
  {\bibfnamefont {P.}~\bibnamefont {Ilten}}, \bibinfo {author} {\bibfnamefont
  {S.}~\bibnamefont {Mrenna}}, \bibinfo {author} {\bibfnamefont
  {S.}~\bibnamefont {Prestel}}, \bibinfo {author} {\bibfnamefont {C.~O.}\
  \bibnamefont {Rasmussen}},\ and\ \bibinfo {author} {\bibfnamefont {P.~Z.}\
  \bibnamefont {Skands}},\ }\bibfield  {title} {\bibinfo {title} {{An
  introduction to PYTHIA 8.2}},\ }\href
  {https://doi.org/10.1016/j.cpc.2015.01.024} {\bibfield  {journal} {\bibinfo
  {journal} {Comput. Phys. Commun.}\ }\textbf {\bibinfo {volume} {191}},\
  \bibinfo {pages} {159} (\bibinfo {year} {2015})},\ \Eprint
  {https://arxiv.org/abs/1410.3012} {arXiv:1410.3012 [hep-ph]} \BibitemShut
  {NoStop}%
\bibitem [{The(2014)}]{TheATLAScollaboration:2014rfk}%
  \BibitemOpen
  \bibfield  {title} {\bibinfo {title} {{ATLAS Pythia 8 tunes to 7 TeV data}},\
  }\href@noop {} {\  (\bibinfo {year} {2014})}\BibitemShut {NoStop}%
\bibitem [{\citenamefont {Buonocore}\ \emph {et~al.}(2021)\citenamefont
  {Buonocore}, \citenamefont {Kallweit}, \citenamefont {Rottoli},\ and\
  \citenamefont {Wiesemann}}]{Buonocore:2021tke}%
  \BibitemOpen
  \bibfield  {author} {\bibinfo {author} {\bibfnamefont {L.}~\bibnamefont
  {Buonocore}}, \bibinfo {author} {\bibfnamefont {S.}~\bibnamefont {Kallweit}},
  \bibinfo {author} {\bibfnamefont {L.}~\bibnamefont {Rottoli}},\ and\ \bibinfo
  {author} {\bibfnamefont {M.}~\bibnamefont {Wiesemann}},\ }\bibfield  {title}
  {\bibinfo {title} {{Linear power corrections for two-body kinematics in the
  $q_T$ subtraction formalism}},\ }\href@noop {} {\  (\bibinfo {year}
  {2021})},\ \Eprint {https://arxiv.org/abs/2111.13661} {arXiv:2111.13661
  [hep-ph]} \BibitemShut {NoStop}%
\bibitem [{\citenamefont {Camarda}\ \emph {et~al.}(2021)\citenamefont
  {Camarda}, \citenamefont {Cieri},\ and\ \citenamefont
  {Ferrera}}]{Camarda:2021jsw}%
  \BibitemOpen
  \bibfield  {author} {\bibinfo {author} {\bibfnamefont {S.}~\bibnamefont
  {Camarda}}, \bibinfo {author} {\bibfnamefont {L.}~\bibnamefont {Cieri}},\
  and\ \bibinfo {author} {\bibfnamefont {G.}~\bibnamefont {Ferrera}},\
  }\bibfield  {title} {\bibinfo {title} {{Fiducial perturbative power
  corrections within the q$_{\bf T}$ subtraction formalism}},\ }\href@noop {}
  {\  (\bibinfo {year} {2021})},\ \Eprint {https://arxiv.org/abs/2111.14509}
  {arXiv:2111.14509 [hep-ph]} \BibitemShut {NoStop}%
\bibitem [{\citenamefont {Catani}\ \emph
  {et~al.}(2019{\natexlab{b}})\citenamefont {Catani}, \citenamefont {Devoto},
  \citenamefont {Grazzini}, \citenamefont {Kallweit},\ and\ \citenamefont
  {Mazzitelli}}]{Catani:2019hip}%
  \BibitemOpen
  \bibfield  {author} {\bibinfo {author} {\bibfnamefont {S.}~\bibnamefont
  {Catani}}, \bibinfo {author} {\bibfnamefont {S.}~\bibnamefont {Devoto}},
  \bibinfo {author} {\bibfnamefont {M.}~\bibnamefont {Grazzini}}, \bibinfo
  {author} {\bibfnamefont {S.}~\bibnamefont {Kallweit}},\ and\ \bibinfo
  {author} {\bibfnamefont {J.}~\bibnamefont {Mazzitelli}},\ }\bibfield  {title}
  {\bibinfo {title} {{Top-quark pair production at the LHC: Fully differential
  QCD predictions at NNLO}},\ }\href {https://doi.org/10.1007/JHEP07(2019)100}
  {\bibfield  {journal} {\bibinfo  {journal} {JHEP}\ }\textbf {\bibinfo
  {volume} {07}},\ \bibinfo {pages} {100}},\ \Eprint
  {https://arxiv.org/abs/1906.06535} {arXiv:1906.06535 [hep-ph]} \BibitemShut
  {NoStop}%
\bibitem [{\citenamefont {Catani}\ \emph {et~al.}(2021)\citenamefont {Catani},
  \citenamefont {Devoto}, \citenamefont {Grazzini}, \citenamefont {Kallweit},\
  and\ \citenamefont {Mazzitelli}}]{Catani:2020kkl}%
  \BibitemOpen
  \bibfield  {author} {\bibinfo {author} {\bibfnamefont {S.}~\bibnamefont
  {Catani}}, \bibinfo {author} {\bibfnamefont {S.}~\bibnamefont {Devoto}},
  \bibinfo {author} {\bibfnamefont {M.}~\bibnamefont {Grazzini}}, \bibinfo
  {author} {\bibfnamefont {S.}~\bibnamefont {Kallweit}},\ and\ \bibinfo
  {author} {\bibfnamefont {J.}~\bibnamefont {Mazzitelli}},\ }\bibfield  {title}
  {\bibinfo {title} {{Bottom-quark production at hadron colliders: fully
  differential predictions in NNLO QCD}},\ }\href
  {https://doi.org/10.1007/JHEP03(2021)029} {\bibfield  {journal} {\bibinfo
  {journal} {JHEP}\ }\textbf {\bibinfo {volume} {03}},\ \bibinfo {pages}
  {029}},\ \Eprint {https://arxiv.org/abs/2010.11906} {arXiv:2010.11906
  [hep-ph]} \BibitemShut {NoStop}%
\bibitem [{\citenamefont {Binoth}\ \emph {et~al.}(2010)\citenamefont {Binoth}
  \emph {et~al.}}]{Binoth:2010xt}%
  \BibitemOpen
  \bibfield  {author} {\bibinfo {author} {\bibfnamefont {T.}~\bibnamefont
  {Binoth}} \emph {et~al.},\ }\bibfield  {title} {\bibinfo {title} {{A Proposal
  for a Standard Interface between Monte Carlo Tools and One-Loop Programs}},\
  }\href {https://doi.org/10.1016/j.cpc.2010.05.016} {\bibfield  {journal}
  {\bibinfo  {journal} {Comput. Phys. Commun.}\ }\textbf {\bibinfo {volume}
  {181}},\ \bibinfo {pages} {1612} (\bibinfo {year} {2010})},\ \Eprint
  {https://arxiv.org/abs/1001.1307} {arXiv:1001.1307 [hep-ph]} \BibitemShut
  {NoStop}%
\end{thebibliography}%

\end{document}